\pdfoutput=1
\documentclass[fleqn,useAMS,usenatbib]{mnras}
\usepackage{newtxtext,newtxmath}
\usepackage[T1]{fontenc}
\usepackage{ae,aecompl}

\usepackage{times}
\usepackage[nodebug]{flushend}
\usepackage{graphicx}
\usepackage{url}
\usepackage{aas_macros}
\hypersetup{pdfauthor={S. J. Murphy},
               pdftitle={WISE J080822.18-644357.3 - a
  45 Myr-old accreting M dwarf hosting a primordial disc},
               bookmarksnumbered=true}

\newcommand\starname{WISE~J0808$-$6443}

\newcommand\masyr{mas\,yr$^{-1}$}
\newcommand\kms{km\,s$^{-1}$}
\newcommand\Msun{\hbox{$M_{\odot}$}}

\newcommand\Lstar{$L_{\star}$}
\newcommand\Rsun{\hbox{$\mathcal{R}^{\rm N}_{\odot}$}}

\newcommand{\logl}{log($L/\mathcal{L}^{\rm N}_{\odot}$)}

\title[45 Myr-old accreting M dwarf]{WISE\,J080822.18$-$644357.3 -- a
  45\,Myr-old accreting M dwarf hosting a primordial disc}

\author[S.~J.~Murphy et al.]{Simon~J.~Murphy\thanks{E-mail:
    s.murphy@adfa.edu.au}$^{1}$, Eric~E.~Mamajek$^{2,3}$ and
  Cameron~P.~M.~Bell$^{4}$\\
$^{1}$School of Physical, Environmental and Mathematical Sciences,
  University of New South Wales Canberra, ACT 2600, Australia\\
$^{2}$Jet Propulsion Laboratory, California Institute of Technology,
  4800 Oak Grove Dr., Pasadena, CA 91109, USA\\
$^{3}$Department of Physics \& Astronomy, University of Rochester,
  Rochester, NY 14627, USA\\
$^{4}$Leibniz Institute for Astrophysics Potsdam (AIP), An der
  Sternwarte 16, 14482 Potsdam, Germany}
\begin{document}

\volume{accepted}
\date{Accepted 2018 February 16. Received 2018 February 13; in original form 2017 March 10}
\label{first page}
\pagerange{\pageref{firstpage}--\pageref{lastpage}}
\pubyear{2018}

\maketitle

\label{firstpage}

\begin{abstract}
WISE~J080822.18--644357.3 (WISE~J0808--6443) was recently identified as a
new M dwarf debris disc system and a candidate member of the
45\,Myr-old Carina association.
Given that the strength of its infrared excess
($L_{\rm{IR}}$/$L_{\rm{\star}}$\,$\simeq$\,0.1) appears to be more
consistent with a young protoplanetary disc, we present the first
optical spectra of the star and reassess its evolutionary and
membership status.
We find \starname\ to be a Li-rich M5 star with strong H$\alpha$
emission ($-125 < \textrm{EW} < -65$\,\AA\ over 4 epochs) whose
strength and broad width are consistent with accretion at a low level
($\sim$10$^{-10}$\,$M_{\odot}$\,yr$^{-1}$) from its disc.
The spectral energy distribution of the star is consistent with a 
 primordial disc and is well-fit using a two-temperature blackbody model
 with $T_{\rm inner}\simeq$ 1100\,K and  $T_{\rm outer}\simeq$ 240\,K.
All\emph{WISE} multi-epoch photometry shows the system exhibits significant variability 
in the 3.4\,\micron\ and 4.6\,\micron\ bands.
We calculate an improved proper motion based on archival astrometry,
and combined with a new radial velocity, the kinematics of the star
are consistent with membership in Carina at
a kinematic distance of 90\,$\pm$\,9\,pc.
The spectroscopic and photometric data are consistent with
\starname\ being a $\sim$0.1\,$M_{\odot}$ Classical T-Tauri star and
 one of the oldest known accreting M-type stars.
These results provide further evidence that the upper limit on the lifetimes of gas-rich
discs -- and hence the timescales to form and evolve protoplanetary systems -- around the lowest mass stars may be longer than previously
recognised, or some mechanism may be responsible for regenerating
short-lived discs at later stages of pre-main sequence evolution.
\end{abstract}


\section{Introduction}
\label{introduction}

Dust and gas-rich discs around pre-main sequence (pre-MS) stars are an inevitable consequence of angular momentum conservation during the star formation process \citep{Shu87,McKee07}.  Discs have been observed around young stars of all masses and can exhibit a variety of morphologies  depending on the myriad external factors (for example, host star mass, multiplicity, dynamical history, proximity to massive stars) and internal processes (accretion, grain coagulation/processing, photo-evaporation, gravitational settling, planet formation) at work throughout their evolution \citep[][and references therein]{Williams11}. 

The most important parameter in understanding disc evolution is the disc lifetime. Firstly, as it reflects the dominant physical processes driving dissipation of the disc but also because it sets an upper limit on the timescale of giant planet formation.  Observationally, the frequency of discs around pre-MS stars rapidly declines with time. Primordial (i.e. first generation) discs supporting ongoing accretion of gas onto the star are common at ages of 1--2~Myr in regions like Orion and Taurus-Auriga, but by 10--15~Myr they appear to be very rare \citep{Haisch01,Hillenbrand05,Sicilia05,Hernandez07b,Mamajek09}.  Near and mid-infrared disc fractions in young clusters and star-forming regions are consistent with median primordial disc lifetimes of 4--5~Myr \citep{Bell13,Pecaut16}, with a similar or slightly shorter lifetime inferred for circumstellar accretion  \citep{Fedele10}.  Variations in disc frequency between supposedly coeval groups can be explained by a variety of factors. For instance, shorter disc lifetimes have been proposed for 
stars in multiple systems \citep[e.g.][]{Bouwman06}, those in the vicinity of massive stars \citep{Guarcello16,Balog07,Johnstone98} and in dense stellar environments \citep[e.g.][]{Pfalzner05,Luhman08,Thies10,Olczak12}.  Conversely, there are several notable examples of gas-rich discs and accretion at ages greater than $\sim$10~Myr.  These include the T Tauri star MP Mus (PDS 66) in the 10--20 Myr-old Lower-Cen-Cru subgroup of Sco-Cen (\citealp{Mamajek02}; but see membership discussion in \citealp{Murphy13}) and a handful of other Sco-Cen accretors \citep{Murphy15,Pecaut16}, the lithium-poor M3+M3 binary St 34 in Taurus \citep{White05} and the A-type stars 49 Ceti \citep[$\sim$40 Myr;][]{Zuckerman12b} and HD 21997 \citep[$\sim$30 Myr;][]{Moor11}.  Long-lived accretion has also been observed outside the immediate solar neighbourhood, including in the $\sim$13~Myr-old double cluster h and $\chi$ Persei \citep{Currie07a,Currie07b}, NGC 6611 \citep{De-Marchi13b}, NGC 3603 \citep{Beccari10} and the Magellanic Clouds \citep{Spezzi12,De-Marchi13a}. 

Unsurprisingly, stellar mass plays a major role in driving disc evolution. Early \emph{Spitzer Space Telescope} surveys of nearby young clusters \citep[e.g.][]{Lada06,Carpenter06,Hernandez07b,Hernandez07,Kennedy09} strongly suggested primordial discs around low-mass stars can persist longer than solar and higher-mass stars.  For example, in their census of discs across the $\sim$11~Myr \citep{Pecaut12} Upper Sco subgroup of Sco-Cen, \citet{Luhman12} \citep[also see][]{Carpenter06,Carpenter09,Chen11} found
10~per cent of members of spectral types B--G possessed inner discs, while this reached $\sim$25~per cent for M5--L0 stars, indicating that a significant fraction of primordial discs around low-mass stars can survive for \emph{at least} 10~Myr.  These findings were recently generalised by \citet{Ribas15}, who undertook an analysis of over 1,400 spectroscopically-confirmed members of nearby ($<$500~pc) associations of ages 1--100~Myr, finding that high-mass ($>$2~\Msun\ in their study) stars dispersed their disks up to twice as fast as low-mass stars.  Together, the results of these studies imply that the mechanisms responsible for removing circumstellar gas and dust operate less efficiently around lower-mass stars and hence longer timescales may be possible for the formation of planetary systems around such hosts.
 
Clearly, the discovery of additional `old' pre-MS stars hosting primordial discs and showing signs of accretion would be useful in better understanding the evolution and dissipation of discs, especially extreme examples much older than 10~Myr.  Through the work of the NASA \emph{Disk Detective} citizen science project\footnote{\url{http://www.diskdetective.org}} \citep[described in][]{Kuchner16}, \cite{Silverberg16} recently identified WISE~J080822.18--644357.3
(hereafter \starname) as an M dwarf 
exhibiting significant excess emission in the 12 and $22\,\rm{\mu m}$ bands of the
\emph{Widefield Infrared Survey Explorer} (\emph{WISE}) survey
\citep{Wright10}. Adopting the proper motion provided by the Southern
Proper Motion Catalog \citep[SPM4;][]{Girard11},
\citeauthor{Silverberg16} used the Bayesian Analysis for
Nearby Young AssociatioNs II tool \citep[\textsc{banyan ii};][]{Malo13,Gagne14} to establish whether the star was associated
with any of the seven nearest young moving groups in the Solar neighbourhood \citep[see reviews by][]{Zuckerman04a,Torres08,Mamajek16}, finding a 93.9 per
cent probability that it is a member of the 45\,Myr-old Carina
association at a distance of $\sim$65\,pc. \citet{Silverberg16} classified the star as hosting
a second-generation debris disc, whereas the strength of its infrared excess
($L_{\rm{IR}}$/$L_{\rm{\star}}$\,$\simeq$\,0.1) appears to be more
consistent with a young primordial disc \citep{Wyatt08}. If such a disc is confirmed, \starname\ would be one of the oldest
known M dwarf protoplanetary disc hosts and a
benchmark object through which to study late-stage disc evolution in
low-mass stars. Despite the proper motion match to Carina,
\citeauthor{Silverberg16} presented no spectroscopic evidence for
either youth or group membership.
In this work we test their claims by
analysing new optical spectra of \starname\ (Section\,\ref{spectra}), deriving a revised proper motion for the star from archival astrometry (Section\,\ref{kinematics}) and reanalysing its
spectral energy distribution, infrared excess (Section\,\ref{disc}) and isochronal age (Section\,\ref{sec:iso}).  With this new information we re-evaluate the evolutionary state
of the star-disc system and discuss its origins with regards to the
dozen or so known young moving groups near the Sun.

\begin{table*} 
\caption[]{Summary of ANU 2.3-m/WiFeS observations of WISE~J080822.18--644357.3.}\label{tab:wifes}
\begin{tabular}{l c c c c c c c c}
\hline
UT Date & Grating & Spectral & RV & EW[\ion{Li}{i}] & EW[H$\alpha$] & $v_{10}$[H$\alpha$] & EW[\ion{He}{i} $\lambda$5876] & EW[\ion{He}{i} $\lambda$6678] \\
 & /Dichroic & Type & (\kms) & (m\AA) & ($\pm$5\,\AA) & ($\pm$10\,\kms) & (\AA) & (\AA) \\ 
\hline
2017 Jan 7 & R7000/RT480 & M4 (veiled?)$^{a}$ & $25.7\pm1.9^{d}$ & $300\pm50$ & $-$110 & 352 & $-$7.5 & $-$2.7 \\
2017 Jan 9 & R7000/RT480 & M4.5$^{a}$ & $22.3\pm1.6^{d}$ & $400\pm30$ & $-$125 & 332 & $-$5.3 & $-$1.8 \\
2017 Feb 6 & R7000/RT480 & M4.5$^{a}$ & $22.5\pm0.5^{d}$ & $390\pm30$ & $-$65 & 298 & $-$6.4 & $-$1.5 \\ 
2017 Feb 6 & R3000/RT560 & M5$^{b}$, M4.8$\pm$0.2$^{c}$ & \dots & \dots & $-$75 & 419 & $-$7.5 & $-$3.3 \\
\hline
\end{tabular}
\begin{flushleft}
$^{a}$ Based on comparison to SDSS M dwarf templates over 5300--7000\,\AA. \\
$^{b}$ Based on comparison to SDSS M dwarf templates over 5500--9200\,\AA. \\
$^{c}$ Average of R1, R2, R3, TiO8465, c81 and VO2 molecular indices, as calibrated by \citet{Riddick07}.\\
$^{d}$ Mean and standard deviation of velocities calculated from 14 (January) and 4 (February) M-type standards.
\end{flushleft}
\label{tab:excesses}
\end{table*}

\section{Spectroscopic observations}  
\label{spectra}

\begin{figure*} 
   \centering
   \includegraphics[width=\linewidth]{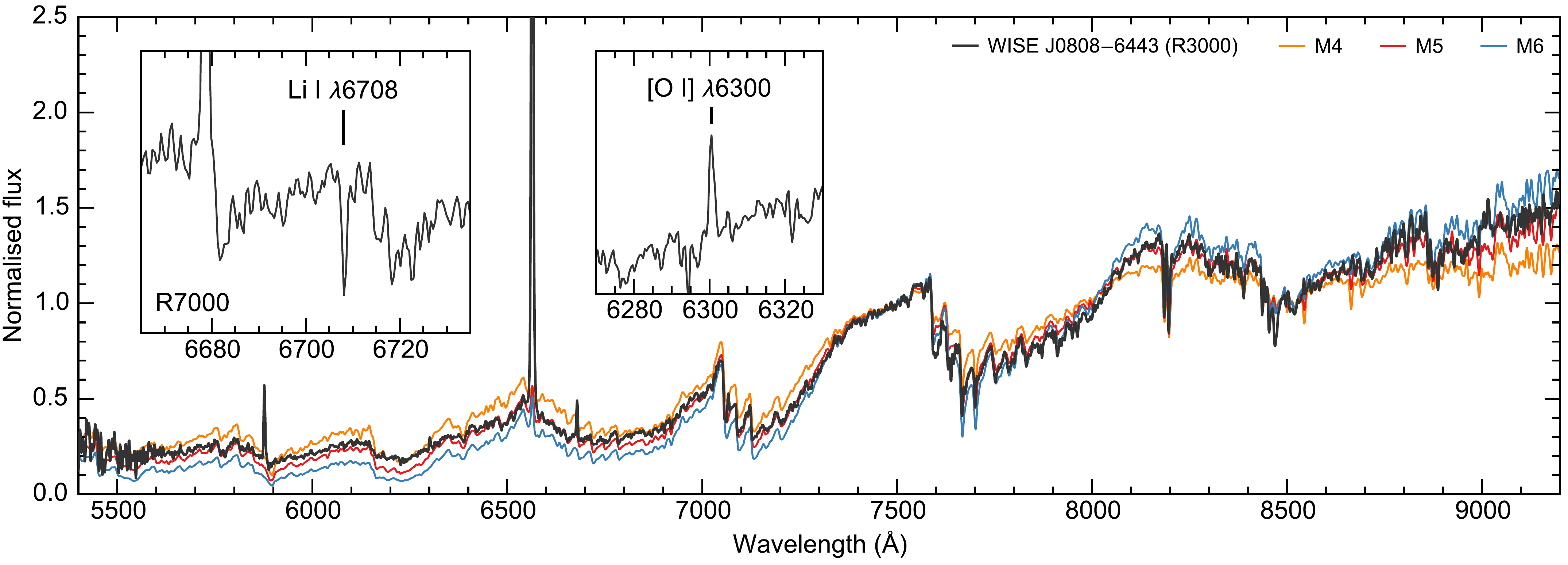}%
   \caption{Telluric-corrected WiFeS/R3000 spectrum of
     \starname\ compared to the SDSS average M dwarf templates of
     \citet{Bochanski07}. Both the WiFeS and SDSS spectra were
     smoothed and normalised at 7500\,\AA\ prior to
     plotting. Strong H$\alpha$, \ion{He}{i} $\lambda$5876 and
     $\lambda$6678 emission is apparent, as well as prominent
     \ion{Na}{i} absorption at 8200~\AA. On this scale the H$\alpha$
     line continues to a peak flux of 4.85 (see also
     Fig.\,\ref{fig:v10}). The insets show the region around the
     \ion{Li}{i} $\lambda$6708 youth indicator and [\ion{O}{i}] $\lambda$6300 emission in 
     the higher-resolution 2017 February 6 R7000 spectrum.  Note
     also the strong \ion{He}{i} $\lambda$6678 emission.}
   \label{fig:spectrum}
\end{figure*}

We acquired four spectra of \starname\, during 2017 January and
February using the Wide Field Spectrograph \citep[WiFeS;][]{Dopita07}
on the ANU 2.3-m telescope at Siding Spring Observatory.  The R7000
grating gives a resolution of $R\approx7000$ over a wavelength range
of 5250--7000\,\AA, while the R3000 grating covers 5400--9700\,\AA\ at
a resolution of $R\approx3000$. Details of the instrument setup and
reduction process, including the measurement of radial velocities, are
provided in \citet{Murphy15} and \citet{Bell17}. The results of these
observations are given in Table~\ref{tab:wifes} and the R3000 spectrum
is shown in Fig.\,\ref{fig:spectrum}.

\starname\ is clearly of mid-M spectral type, with pronounced
H$\alpha$ and \ion{He}{i} emission. We also detected strong
\ion{Li}{i} $\lambda$6708 absorption in the R7000 spectra and weak
($\sim$1\,\AA) forbidden [\ion{O}{i}]\,$\lambda$6300 emission in all
four observations.  We measured the equivalent width (EW) of the broad
H$\alpha$ line by direct integration of the line profile. After
comparing the R3000 spectrum to the SDSS average M dwarf templates of
\citet{Bochanski07} we estimate a spectral type of approximately M5,
in agreement with the M4.8$\pm$0.2 average of the R1, R2, R3, TiO8465,
c81 and VO2 molecular indices (covering 7800--8500\,\AA), as
calibrated by \citet{Riddick07}. From the three R7000 spectra we
calculate a weighted mean radial velocity of $22.7\,\pm\,0.5$\,\kms\ and
$\textrm{EW[\ion{Li}{i}]} = 380\,\pm\,20$\,m\AA. Within the limits of
the modest WiFeS resolution, the radial velocity appears
constant on timescales of $\sim$1 month and the star's cross
correlation function is consistent with a slowly-rotating ($v\sin i
\lesssim 45$~\kms) single star \citep[see discussion in][]{Murphy15}.

\subsection{H$\alpha$ emission and accretion} 
\label{accretion}

\begin{figure}  
   \centering
   \includegraphics[width=\linewidth]{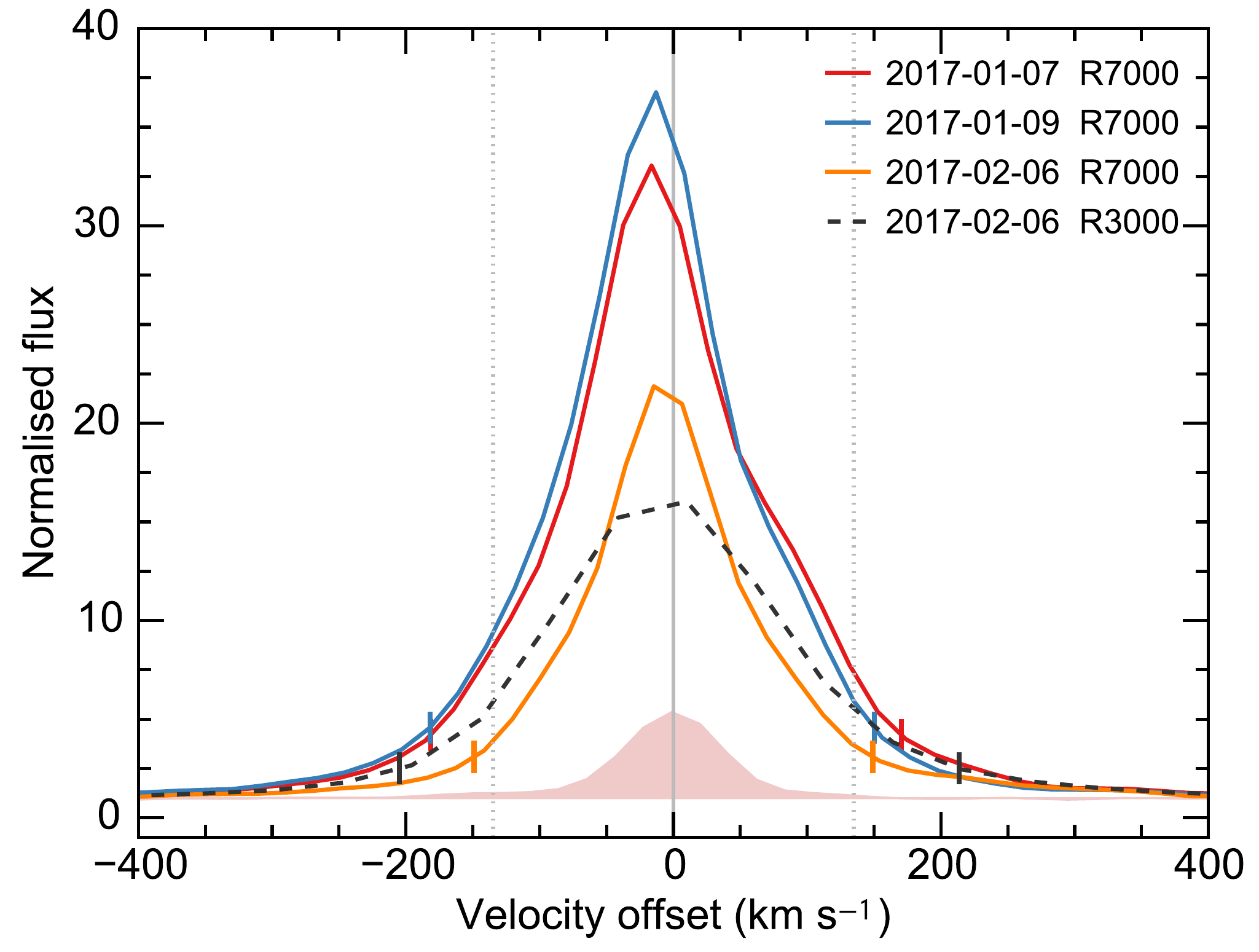}%
   \caption{WiFeS H$\alpha$ velocity profiles of \starname.  All
     spectra have been shifted to the heliocentric rest
     frame. Vertical markers show the velocities at which the flux is
     10 per cent of the peak value. The two dotted lines are the
     (symmetric) $v_{10}=270$\,\kms\ accretion criterion of
     \citet{White03}. For comparison, the shaded region shows the
     H$\alpha$ profile of the new M5 TW Hydrae association member and non-accretor TWA~36 \citep*[$\textrm{EW} = -9.5$\,\AA, $v_{10} =
       169$~\kms;][]{Murphy15b}.}
   \label{fig:v10}
\end{figure}

\begin{figure} 
   \centering
   \includegraphics[width=\linewidth]{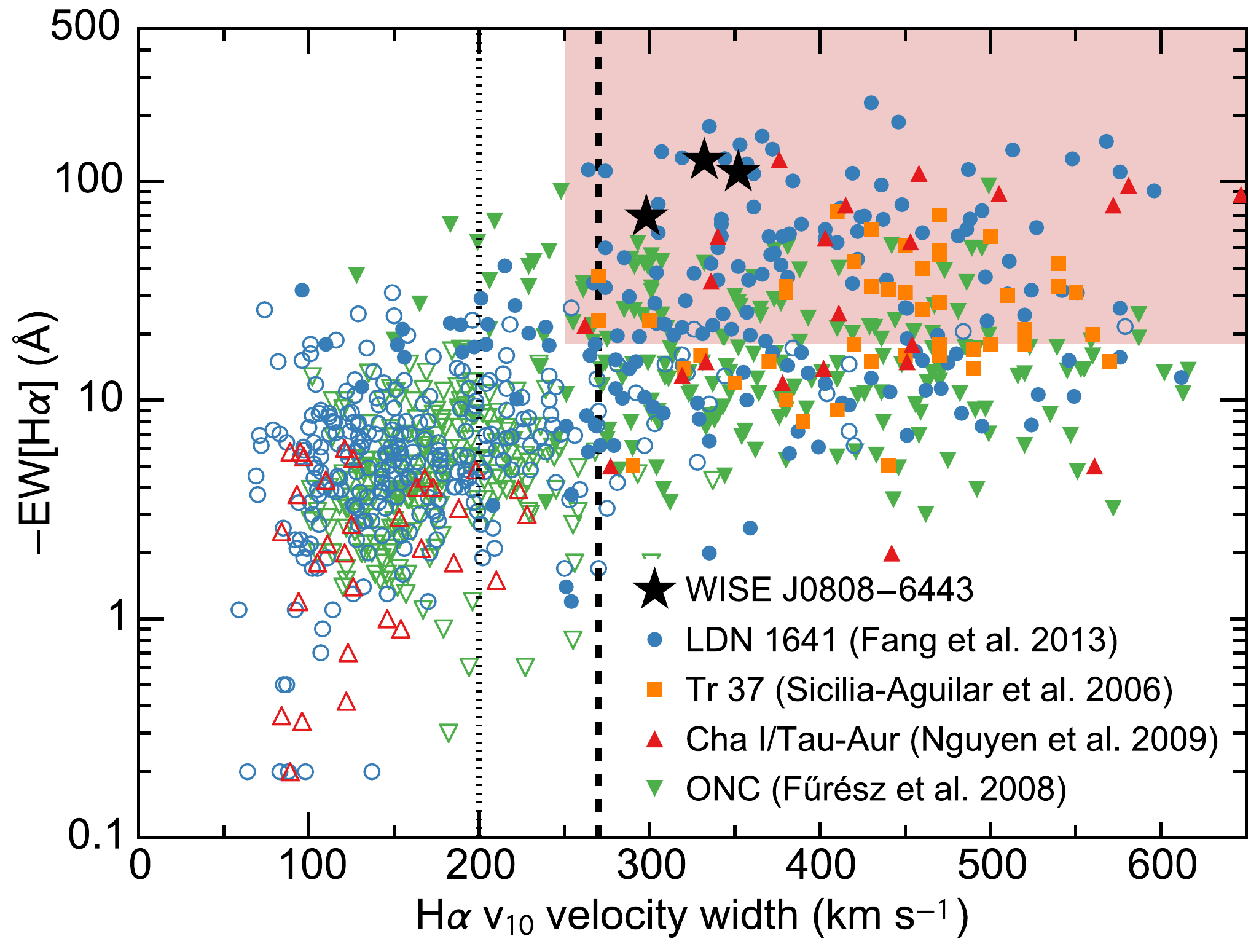}%
   \caption{EW[H$\alpha$] and $v_{10}$[H$\alpha$] measurements for
     \starname\ compared to members of young clusters and
     star-forming regions; LDN 1641 \citep{Fang13},
     Trumpler 37 \citep{Sicilia06b}, Cha I and
     Taurus-Auriga \citep{Nguyen09} and the Orion Nebula
     Cluster \citep{Furesz08}. Filled points
     are accretors with confirmed inner discs from \emph{Spitzer},
      open points are non-accretors. The dotted and dashed lines
     mark the $v_{10}$ accretion criteria
     proposed by \citet{Jayawardhana03} and \citet{White03},
     respectively, while the shaded region shows the criteria for an M5
     accretor proposed by \citet{Fang09,Fang13}.}
   \label{fig:ctts}
\end{figure}

Accretion of gas-rich material from the inner regions of a
circumstellar disc onto the star is typically accompanied by enhanced
Balmer and other line emission \citep[e.g.][]{Muzerolle98}. The
H$\alpha$ emission observed in \starname\ is much larger than
typically seen in mid-M field stars and immediately suggests the star
is accreting.  \citet{White03} and \citet{Barrado03} found that mid-M
stars with $\textrm{EW[H$\alpha$]}<-20$\,\AA\ tend to be T-Tauri stars
with other corroborating evidence of accretion (strong IR excess,
veiling). Similarly, by comparing the \emph{Spitzer} infrared colours
of young stars with and without dusty inner discs, \citet{Fang09}
proposed an EW limit of $-$18\,\AA\ for an M5 accretor.  Over three
nights we measured EW[H$\alpha$] values in the range $-125 <
\textrm{EW} < -65$\,\AA, considerably in excess of these criteria (see Fig.\,\ref{fig:ctts}).

The width of the H$\alpha$ line (typically quantified by the
velocity full width at 10 per cent of peak flux, or $v_{10}$) is also
used to distinguish accretors from stars whose narrower emission
is due to chromospheric activity.  \citet{White03} proposed a limit of
$v_{10}>270$~\kms\ for accretors independent of spectral type, while
\citet{Jayawardhana03} and \cite{Fang13} have suggested limits of
200\,\kms\ and 250\,\kms, respectively.  Fig.\,\ref{fig:v10} shows the
H$\alpha$ line profiles of our four WiFeS spectra. We calculated
$v_{10}$ values by fitting a linear function over
$\pm$500--1000\,\kms\ and normalising the spectra, then determining
the two velocities at which the flux profile fell to one tenth of its
peak value (see Fig.\,\ref{fig:v10}). These $v_{10}$ measurements are
given in Table\,\ref{tab:wifes} and Fig.\,\ref{fig:ctts}, and are
clearly indicative of accretion.

The variability of the EW and $v_{10}$ values in
Table\,\ref{tab:wifes} is commonly seen in other accretors
\citep[e.g.][]{Nguyen09,Costigan12} and probably indicates variability
in the accretion rate. Applying the $v_{10}$[H$\alpha$]--$\dot{M}_{\rm acc}$ relation of
\citet{Natta04} to our R7000 spectra yields mass accretion rates of
(1.0--3.3)\,$\times10^{-10}$\,$M_{\odot}$\,yr$^{-1}$
($-10<\log\dot{M}_{\rm acc}/M_{\odot}\,\textrm{yr}^{-1}<-9.5$), with
an uncertainty of 0.4\,dex
(2.5$\times$).  These rates are
comparable to those observed in low-mass members of the TW Hydrae
association, $\eta$ Cha cluster and Scorpius-Centaurus OB association
\citep{Muzerolle00,Lawson04,Murphy15b} and are 1--2
orders of magnitude below those typically inferred in young
star-forming regions such as Taurus and $\rho$ Ophiuchus
\citep[e.g.][]{Calvet00}.  We note that the \citet{Natta04} relation was
determined from a sample of few Myr-old stars and is not strictly
applicable to older pre-MS stars, whose radii will be smaller and
surface gravities higher. For example, assuming a constant accretion luminosity
$L_{\rm acc}$, the radius and hence mass accretion rate $\dot{M}_{\rm
acc} \propto L_{\rm acc} R_{\star} / G M_{\star}$ onto a
0.1$\,M_{\odot}$ star is $\sim$4$\times$ (0.6 dex) smaller at an age
of 40\,Myr compared to 2\,Myr \citep{Feiden16}. The accretion rates derived above do not include any corrections for age.
 
\subsection{Lithium depletion and low-gravity features}

\begin{figure}  
   \centering
   \includegraphics[width=\linewidth]{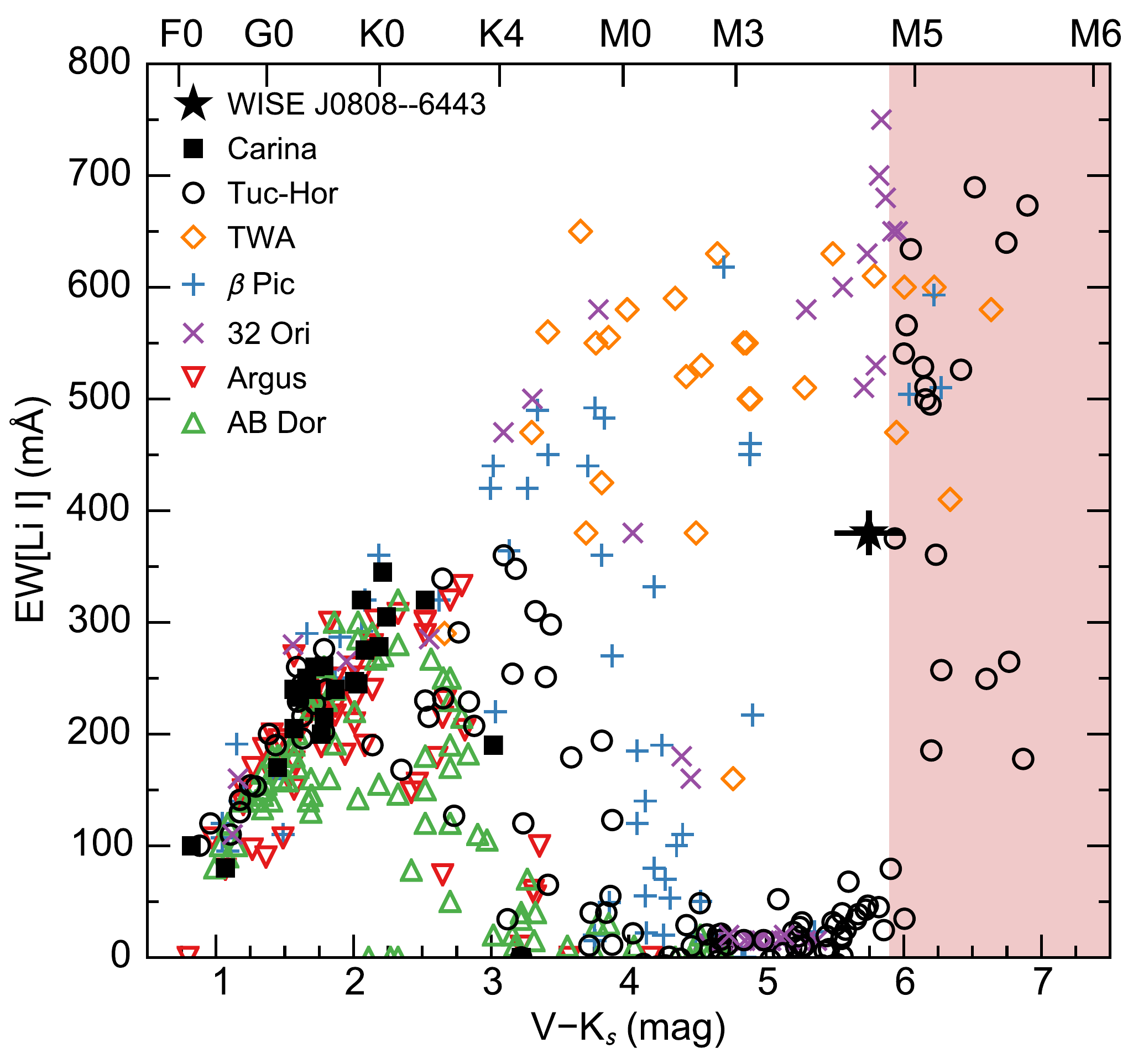}%
   \caption{EW[\ion{Li}{i}] of \starname\ compared to
     other young moving group members, compiled from the studies
     of \protect\cite{daSilva09}, \protect\citet{Schneider12a}, \protect\cite{Kraus14},
     \protect\cite{Malo14b}, \protect\cite{Binks16} and
     \citet{Bell17}. For \starname\ we assume an uncertainty of 0.2\,mag on
     its faint SPM4 $V$-band magnitude. The shaded region denotes the Li-rich side of the Tuc-Hor lithium depletion boundary.
     }
   \label{fig:li}
\end{figure}

Because lithium is easily destroyed in stellar interiors, with some caveats
the presence of \ion{Li}{i} $\lambda$6708 absorption in a star is a
sign of youth.  As young stars contract toward the
main-sequence, their core temperatures rise until at $\sim$$3 \times
10^{6}$\,K lithium burns. These temperatures can be reached in either
fully-convective mid- to late-M dwarfs or at the base of the
convective zone in late-K or early-M dwarfs. Between these luminosities,
rapid depletion ensues and the \ion{Li}{i} $\lambda$6708 feature is no
longer visible. This mass-dependent behaviour leads to the depletion
patterns seen in Fig.\,\ref{fig:li}, where we plot the EW[Li] of
\starname\ compared to young moving group members from the literature.
Based on the moderate amount of depletion observed in the star, we can
immediately infer that it is older than the $\sim$10\,Myr TW Hydrae
association (TWA), whose mid-M members are essentially undepleted.

The Carina association is believed to be part of a larger complex
\citep[the so-called Great Austral Young Association or
  GAYA;][]{Torres01,Torres08} comprising itself and the Columba and
Tucana-Horologium associations. Although spatially and kinematically
distinct, \cite{Bell15} found that all three associations are
co-eval, with ages of $\sim$45\,Myr. Unlike Carina, which is a
particularly sparse association of a few dozen high-probability
members (see Section\,\ref{membership}), Tuc-Hor contains hundreds of
stars \citep{Kraus14}. This census provides a well-sampled depletion
pattern for stars at age $\sim$45\,Myr and we may use this to assess
whether the Li depletion of \starname\ is consistent with such an
age. From Fig.\,\ref{fig:li} it is clear that \starname\ occupies a
region between the Li-poor early M-type Tuc-Hor members and the
Li-rich mid to late M-type members. This sharp discontinuity (see the
shaded region of Fig.\,\ref{fig:li}) defines the lithium depletion
boundary (LDB) of the association and its luminosity provides a
precise, almost model-independent age of $40\pm3$\,Myr
\citep{Kraus14}. The position of \starname\ in Fig.\,\ref{fig:li} is
entirely consistent with the expected LDB position for a co-eval
population of nominal age 40--45\,Myr. A strong upper age limit is
provided by the 80--90\,Myr LDB age of the $\alpha$ Persei cluster,
whose LDB occurs at a spectral type of M6--6.5
\citep{Stauffer99,Barrado04b}.
 
\begin{figure}  
   \centering \includegraphics[width=\linewidth]{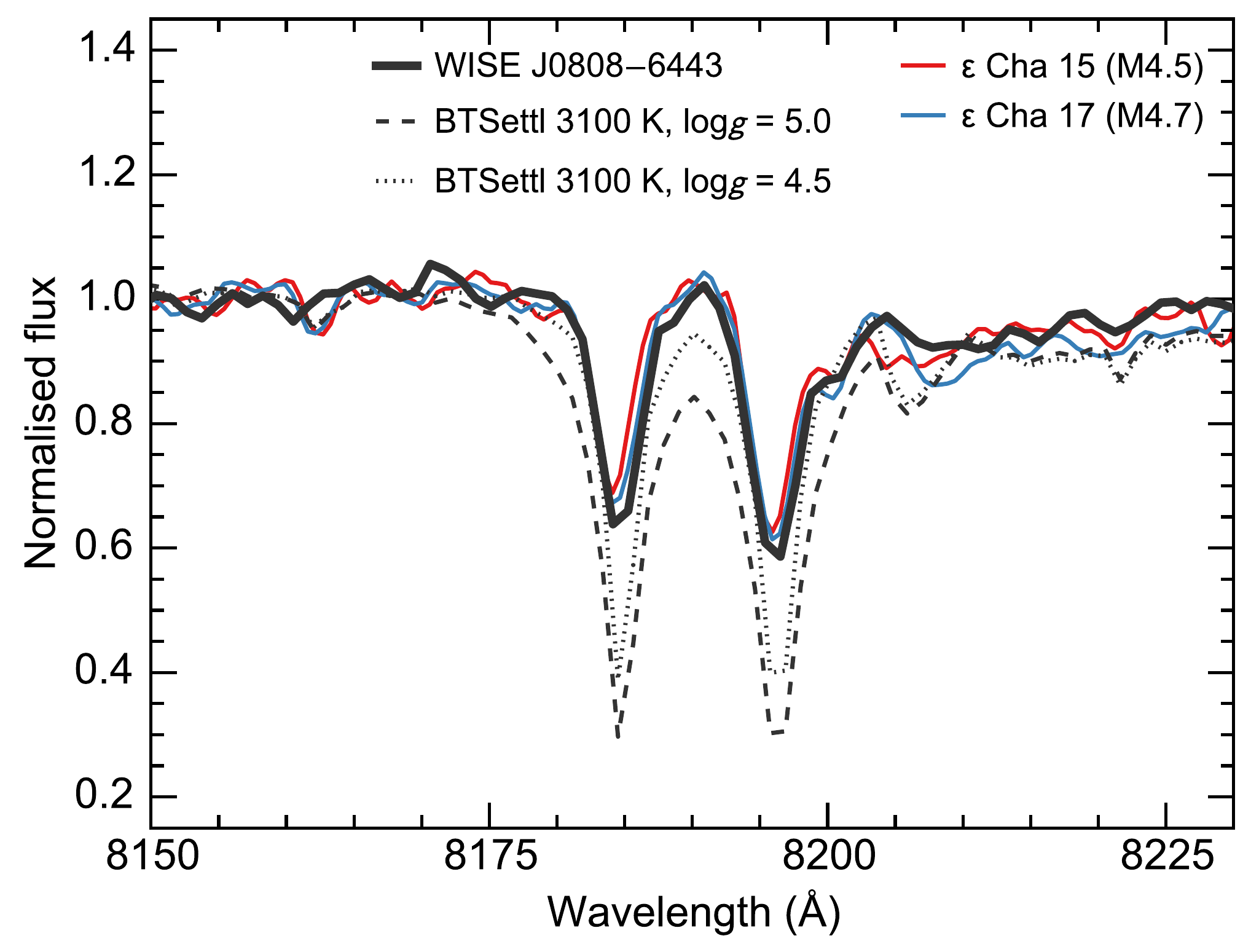}%
   \caption{WiFeS/R3000
   spectrum around the 8200\,\AA\ \ion{Na}{i} doublet. The strength
   of \ion{Na}{i} absorption is similar to members of the young
   $\epsilon$ Cha association \citep{Murphy13}. Also plotted are two
   3100\,K BT-Settl-CIFIST model atmospheres \citep{Baraffe15}, smoothed to
   $R\approx3000$ and placed on the same wavelength scale
   as \starname. The absorption in \starname\ is much weaker than the
   $\log g=5.0$ model, particularly in the broad line wings, but
   similar to the $\log g=4.5$ spectrum. } \label{fig:Na}
\end{figure}

As the radius of a star decreases during its 
pre-MS evolution, surface gravity-sensitive absorption features, such as those
of the alkali metals or metal hydrides, can be used as indicators of a
star's youth. At mid-M spectral types, one of the best gravity-sensitive
features is the \ion{Na}{i} doublet at $\lambda$8183/8195
\citep{Schlieder12}. While the \ion{Na}{i} line strength can not provide a precise estimate of age, it has been widely used
to distinguish young cluster members from giants and Gyr-old field
stars \citep[e.g.][]{Slesnick06}. We plot in Fig.\,\ref{fig:Na} the
region around the \ion{Na}{i} doublet for \starname\ and two members
of the 3--5~Myr-old $\epsilon$~Cha association of similar spectral
type, observed with the same instrument settings
\citep{Murphy13}. The strength of the absorption in
\starname\ ($\textrm{EW[\ion{Na}{i}]}\approx4$\,\AA) is very similar to the two young
stars, and is much weaker than a dwarf model atmosphere of similar
temperature. This shows that \starname\ has a weaker surface
gravity and hence younger age than a field M dwarf. 

In light of its strong H$\alpha$ emission, lithium absorption and low
gravity, we spectroscopically classify \starname\ as an accreting,
Classical T-Tauri Star (CTTS).

\section{Proper motion and kinematics}
\label{kinematics}

\citet{Silverberg16} assigned \starname\ as a candidate member of the
Carina association based entirely on its position and SPM4 proper
motion, considering no other astrometric catalogues.  The SPM4 proper
motion has large uncertainties ($\sim$7\,mas\,yr$^{-1}$) and so a
critical examination of other proper motion measurements and an
independent estimate is warranted.  Table~\ref{tab:proper_motion}
lists the published proper motions for \starname\ and our new
measurement (see below). Notably, the star does not appear in any UCAC
catalogue (including UCAC5).
 It is apparent that the proper motion of \starname\ is not well constrained, with large differences between
catalogues
\citep[e.g. $\Delta\mu_{\delta}\approx19$\,mas\,yr$^{-1}$ between 
  SPM4 and PPMXL;][]{Roeser10}.  Furthermore, the USNO B1.0
catalogue \citep{Monet03} lists a proper motion of zero for this star,
which signifies it was unable to detect a statistically significant
proper motion. Given the ambiguity with regards to its proper motion,
to assign membership of \starname\ -- and by extension an age -- on
the basis of a single poorly-constrained observable is clearly
insufficient.

We calculated a new proper motion for \starname\ using the published
positions compiled in Table~\ref{tab:astrometry} and the least-squares
formulae of \citet{Teixeira00}. Given the heterogeneity of the source
catalogues and heteroskedasticity of their uncertainties, we assessed
the proper motion uncertainties by multiplying the least-squares
estimates by their Birge ratios \citep{Mohr05}, in effect forcing the
reduced $\chi^2$ of the linear astrometric fits to be $\sim$1.  As
seen in Table~\ref{tab:proper_motion}, the differences in proper
motion uncertainties between the simple least squares estimates and
the Birge ratio-adjusted estimates are negligible. We adopt the latter
for the kinematic calculations described below. Our new proper motion
agrees with SPM4 to within the latter's larger uncertainties and is
similar to values recently published in the HSOY
catalogue \citep[][see Table\,\ref{tab:proper_motion}]{Altmann17},
which was formed by combining \emph{Gaia} DR1
astrometry \citep{Gaia16} with PPMXL.

\begin{table}
\caption[]{Estimated proper motions for WISE~J080822.18--644357.3.}
\begin{tabular}{l c c}
\hline
Reference & $\mu_{\alpha}\cos\delta$   & $\mu_{\delta}$\\
       & (mas\,yr$^{-1}$)& (mas\,yr$^{-1}$)\\
\hline
SuperCOSMOS \citep{Hambly01} & $-6.8\pm4.8$ & $18.5\pm4.9$\\
USNO-B1.0 \citep{Monet03} & $0\pm0$ & $0\pm0$\\
PPMXL \citep{Roeser10} & $-11.2\pm8.5$ & $17.8\pm8.5$\\
SPM4$^{a}$ \citep{Girard11} & $-9.9\pm7.1$ & $36.7\pm7.1$\\
All\emph{WISE} \citep{Cutri14} & $53\pm27$ & $-14\pm27$\\
HSOY \citep{Altmann17} & $-14.0\pm3.1$ & $27.9\pm3.1$\\
This work$^{b}$ & $-12.5\pm1.9$ & $29.4\pm2.2$\\
This work (adopted)$^{c}$ & $-12.5\pm2.1$ & $29.4\pm2.5$\\
\hline
\end{tabular}
\begin{flushleft}
$^{a}$ Value adopted by \citet{Silverberg16}.\\  
$^{b}$ Based on a simple least-squares fit.\\
$^{c}$ Calculated as in $^{b}$, but multiplies the uncertainty by a
  ratio taking into account the $\chi^{2}$ of the fit.\\
\end{flushleft}
\label{tab:proper_motion}
\end{table}

\begin{table*}
\caption[]{Astrometry used to calculate the proper motion of WISE~J080822.18--644357.3.}
\begin{tabular}{c c c c c l}
\hline
$\alpha$   &   $\delta$   &   $\sigma_{\alpha}$   &   $\sigma_{\delta}$   &   Epoch   &   Reference\\
(deg.)   &   (deg.)   &   (mas)   &   (mas)   &   (yr or JD)   &   \\
\hline
122.0924291146 & $-$64.7325757593 & 0.407 & 0.687 & 2015.0    & \emph{Gaia} DR1 \citep{Gaia16}\\
122.0924381    & $-$64.7326092    & 34    & 35.8  & 2010.5589 & All\emph{WISE} \citep{Cutri14}\\
122.092536     & $-$64.732738     & 400   & 400   & 2450916.764023 &   DENIS DR3 \citep{Epchtein99}\\
122.092519     & $-$64.732778     & 400   & 400   & 2451237.607512 &   DENIS DR3 \citep{Epchtein99}\\
122.0925957    & $-$64.7327013    & 69.5  & 69.5  & 2000.0    & SPM4 \citep{Girard11}\\
122.092535     & $-$64.732719     & 330   & 340   & 1991.131  & GSC2.3.2 \citep{Lasker08}\\
122.092670     & $-$64.732823     & 66    & 83    & 1985.1    & USNO-B1.0 \citep{Monet03}\\
\hline
\end{tabular}
\label{tab:astrometry}
\end{table*}

\subsection{Young moving group membership}
\label{membership}

\begin{figure*}   
   \centering
   \includegraphics[width=\textwidth]{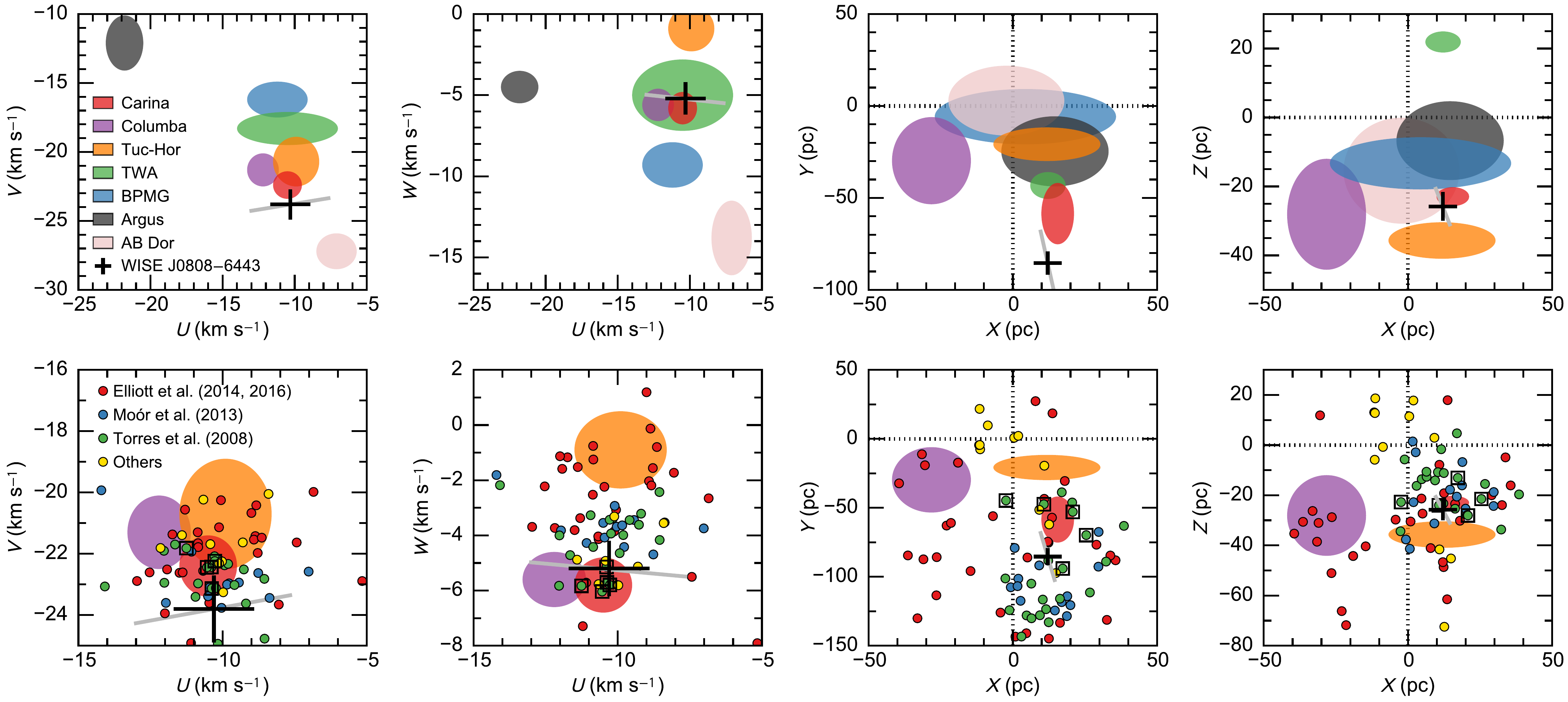}%
   \caption{\emph{Top row:} Three-dimensional Galactic velocity and
     position of \starname\ relative to several nearby, young moving
     groups from \citet{Malo14a}. Uncertainties for \starname\ were
     calculated from uncertainties on the proper motion, radial
     velocity and distance. The grey line shows the change in velocity
     and position as the $90\pm9$\,pc assumed distance changes by
     $\pm2\sigma$. Based on this comparison, \starname\ is a likely
     Carina member. \emph{Bottom row:} As above, but only showing the
     Carina, Columba and Tuc-Hor associations. Points are proposed
     Carina members from the literature with trigonometric parallaxes,
     predominantly from \emph{Gaia} DR1. The five square markers are
     the `bona fide' members of \citet{Malo13,Malo14a}. These stars
     were the only \citet{Torres08} members with \emph{Hipparcos}
     parallaxes.}
   \label{fig:uvwxyz}
\end{figure*}

With our revised proper motion and radial velocity measurement we
can better test the membership of \starname\ in Carina or another of the
 young moving groups which populate the Solar neighbourhood. The
\textsc{banyan ii} Bayesian membership tool (v1.4) now returns a
membership probability for Carina of 91.7\,per cent (assuming
$<$\,1\,Gyr age priors) at an inferred distance of $83\pm7$\,pc. This
is comparable to the 93.9 per cent probability reported by
\citeauthor{Silverberg16} at $65_{-8}^{+9}$\,pc using just the SPM4 proper
motion.  No other young group tested by \textsc{banyan} (TW Hya,
$\beta$~Pic, Columba, Tuc-Hor, Argus, AB Dor) returned a non-zero
membership probability, with the balance of probabilities going to the
young field hypothesis.  
To investigate the robustness of the \textsc{banyan} results, 
we also tested the literature proper motions from Table\,\ref{tab:proper_motion}.  All returned most-likely memberships in either 
Carina (SPM4: 98\,per\,cent, HSOY: 93\,per\,cent) or the young field population 
(SuperCOSMOS, AllWISE, PPMXL). In the latter case, the balance of 
membership probabilities always went to Carina ($P_{\textrm{Car}}=1.5, 3.8$ and 32.5\,per\,cent, respectively). No other group returned a non-zero probability.

Several nearby, young moving groups are not
included in \textsc{banyan} -- we additionally tested membership of
\starname\ in $\eta$ and $\epsilon$ Cha \citep[including ejection
  from $\eta$ Cha;][]{Murphy10}, Octans and the Lower Cen-Cru 
subgroup of the Sco-Cen OB association, and could not find a
satisfactory kinematic or spatial match (e.g. low membership probabilities,
high peculiar motions, predicted radial velocities which disagree
with measured value). \starname\ is also 
in the vicinity of the IC 2602 and IC 2391 open clusters, but neither
provide a plausible kinematic fit. Carina appears to be the only known
 group to which \starname\, could plausibly belong\footnote{Since this work was accepted for publication, \textsc{banyan ii} has been replaced by \textsc{banyan $\Sigma$} \citep{Gagne18}, which provides several improvements including spatio-kinematic models for 21 additional young groups aged 1--800 Myr. \starname\  now returns a Carina membership probability of 89 per~cent at an inferred distance of $88\pm6$~pc, in good agreement with our results above. No other young group returned a non-zero membership probability, with the balance of probabilities going to the field model. The \textsc{banyan $\Sigma$} kinematic model for Carina contains seven stars -- the original membership list of \cite{Malo14a} in addition to the new K3 member HD 298936 \citep[TWA 21; see][]{Gagne17}.}.

Adopting the \citet{Malo14a} mean space motion for Carina and our new
proper motion for \starname, we use the convergent point method
\citep[e.g.][]{Mamajek05} to calculate a kinematic parallax
($\varpi_{\rm{kin}}$ = 11.16\,$\pm$\,1.13 mas; $d$ = 90\,$\pm$\,9 pc)
and predicted radial velocity (21.5\,$\pm$\,1.2 km\,s$^{-1}$).  The
larger distance stems from the revised proper motion having a smaller
magnitude (32\,\masyr) than the SPM4 value (38\,\masyr).  The star's
proper motion is pointing towards Carina's convergent point, with only
2.4\,$\pm$\,2.2\,\masyr\ of peculiar motion. We calculate a
heliocentric position of $(X,Y,Z) = (12.1, -85.4, -25.8)\pm(4.9, 6.3,
4.2)$\,pc and a space motion of $(U,V,W)=(-10.3, -23.8, -5.2)\pm(1.4,
1.1, 1.0)$\,\kms, which agrees with the \citet{Malo14a} mean group
velocity at better than 2\,\kms.

\subsection{Carina members in the literature}

In the top row of Fig.\,\ref{fig:uvwxyz} we plot the position
and velocity of \starname\, against the mean values for young
moving groups and associations within 100\,pc from \citet{Malo14a}. While
there is excellent spatial and kinematic agreement between
\starname\ and Carina in this diagram, it is important to note that
their Carina model is based on only five stars; AB Pic, HD 49855, HD
55279, V479 Car, and HD 83096AB, these being the only stars in the
original membership list of \citet{Torres08} with \emph{Hipparcos}
parallaxes. The updated models of \citet{Gagne14} used by
\textsc{banyan ii} excluded V479 Car and added the nearby M dwarfs GJ 2079
and GJ 1167 \citep{Shkolnik12}, but with only six systems the mean
properties of Carina are poorly defined.  We have gathered
proposed Carina members from the literature with radial velocities and
parallaxes \citep[predominantly new measurements from \emph{Gaia}
  DR1;][]{Gaia16} and computed their space motions and positions which are
plotted in the bottom row of Fig.\,\ref{fig:uvwxyz}. These candidates came
mainly from the memberships of \citet{Torres08}, \citet{Moor13} and 
\citet{Elliott14,Elliott16}, with additional stars
from \citet{Viana09}, \citet{Shkolnik12}, \citet{Riedel14} and
\citet{Bowler15}. Many
of these stars have also been proposed by various authors 
as members of other young moving groups , especially Columba, with which
Carina shares a similar position and velocity (see
Fig.\,\ref{fig:uvwxyz}). For example,
\citet{Elliott14} reclassify seven \citet{Torres08} members (including
all five of the \emph{Hipparcos} stars above) as either Columba
members or Carina non-members, although their methodology or rationale
is not given.

The general distribution of Carina candidates in
Fig.\,\ref{fig:uvwxyz} is spatially and kinematically distinct from
both Columba and Tuc-Hor (whose mean phase space positions are much
better defined from several tens of members with parallaxes). Carina
is clearly much larger than defined by the five \emph{Hipparcos}
stars \citep[which do not appear to be Columba members,
c.f.][]{Elliott14}, especially in the negative $Y$ direction. This is
similar to the solution proposed by
\citet{Torres08} with kinematic distances and is understandable as the limited depth of
\emph{Hipparcos} would favour nearby stars. Assuming they are all bona fide Carina members, from the 74 stars with parallaxes and radial velocities we calculate a mean space motion of $(-10.4,-22.2,-3.5)\pm(1.8,1.4,1.9)$\,\kms\ (1$\sigma$ variation), which is 2.3\,\kms\ larger in $W$ than the \citet{Malo14a} velocity.  If true, the revised space motion does not change our derived 90\,pc kinematic distance by more than a parsec. 

There are several proposed Carina members in the immediate vicinity of
\starname. The \citet{Moor13} star TYC 8933-1204-1 is only
1.40$^{\circ}$ away and, while it does not possess a \emph{Gaia} DR1
parallax, its UCAC5 proper motion is only 5\,\masyr\ from that of
\starname\ and the stars' radial velocities agree to 1~\kms.  Of
the proposed members with parallaxes, the G5 star
TYC~8929-927-1 \citep{Torres08} is only 2.07$^{\circ}$ away, with radial
velocity $23.4\pm0.5$\,\kms \citep{Torres06} and $\varpi = 10.85\pm0.31$\,mas ($d = 92.2\pm2.6$\,pc).  The
star's \emph{Gaia} proper motion matches that of \starname\, to within
$\sim$1\,\masyr.  With a projected separation of
$\sim$3.3\,pc, it is unlikely that they are bound, but more likely
unbound members of the same young moving group. Given their congruent 
proper motions, the \emph{Gaia} parallax distance
for TYC~8929-927-1 corroborates our kinematic distance of
90\,$\pm$\,9 pc for \starname.  Finally, we note that the
\cite{Gagne15b} Carina candidate 2MASS\,J08045433$-$6346180 ($<$M2; 99 per cent \textsc{banyan} membership probability) is only 1.03$^{\circ}$
from \starname. Aside from being X-ray bright
(1RXS\,J080455.0$-$634621), little is known about this star, though
given its much larger UCAC5 proper motion it cannot be co-distant with
\starname.

While a critical re-evaluation of all proposed Carina members in light
of new \emph{Gaia} parallaxes is beyond the scope of this work, given
the close spatial and kinematic agreement between \starname\, and the
stars in Fig.\,\ref{fig:uvwxyz}, as well as the presence of other
proposed Carina members in its vicinity, we see no reason not to
assign membership to the association. Unambiguous confirmation,
however, will require a refined Carina membership and a trigonometric
parallax.  \emph{Gaia} should provide a parallax for a $G=16$~mag M5
star with an estimated end-of-mission uncertainty of approximately
45\,$\mu$as\footnote{\url{http://www.cosmos.esa.int/web/gaia/science-performance}}.

\section{Circumstellar disc emission}
\label{disc}

\citet{Silverberg16} noted that \starname\ has a strong infrared
excess in the \emph{WISE} $W3$ (12\,$\mu$m) and $W4$ (22\,$\mu$m)
bands and fitted the star's spectral energy distribution (SED) with a
2900\,K stellar atmosphere and $\sim$260\,K blackbody of luminosity
$L_{\rm IR}/L_{\star}=0.081$. In this section we re-examine the
infrared excess using models and photometry provided by the Virtual
Observatory SED Analyzer \citep[VOSA v5.1;][]{Bayo08}.

Within VOSA we first gathered all available photometry for
\starname\ from photometric catalogues. The resulting SED contained
fluxes from \emph{Gaia} DR1 ($G$), DENIS
\citep[$iJK$;][]{Epchtein99}, 2MASS \citep[$JHK_{s}$;][]{Cutri03} and
All\emph{WISE} \citep[$W1$--$W4$;][]{Cutri14}. Following
\citeauthor{Silverberg16} we fit synthetic photometry derived from
solar-metallicity BT-Settl-CIFIST models \citep[$\Delta T_{\rm
    eff}=100$\,K;][]{Baraffe15,Allard12}, restricting the surface gravity to
$\log g= 4.5$. The excess finding routine in VOSA identified a
possible excess redward of $W1$ so in fitting the models we did not
include the four \emph{WISE} bands or the broad \emph{Gaia} $G$-band,
which may be contaminated by the accretion-driven strong H$\alpha$
line or any blue continuum excess\footnote{By including the $W1$
  magnitude in the VOSA BT-Settl model fit we were able to replicate
  the \citeauthor{Silverberg16} results and fit a 2900\,K model
  atmosphere to the $(iJK)_{\rm DENIS}JHK_{s}W1$ photometry. However,
  the $\chi^2$ of this fit was much larger than the 3100\,K fit
  obtained in this study.}. For this reason we also excluded the SPM4
$B$- (photographic) and $V$-band photometry.  Furthermore, we do not
consider the effects of reddening in this analysis -- at a kinematic
distance of 80--90\,pc, \starname\ is expected to lie within the Local
Bubble and reddening maps predict its reddening and extinction should
be negligible \citep[$E(b-y)< 0.01$, $A_V<0.04\,\rm{mag}$;][]{Reis11},
although we cannot exclude the possibility of residual reddening  along the line of sight.

\begin{figure}  
   \centering
   \includegraphics[width=\linewidth]{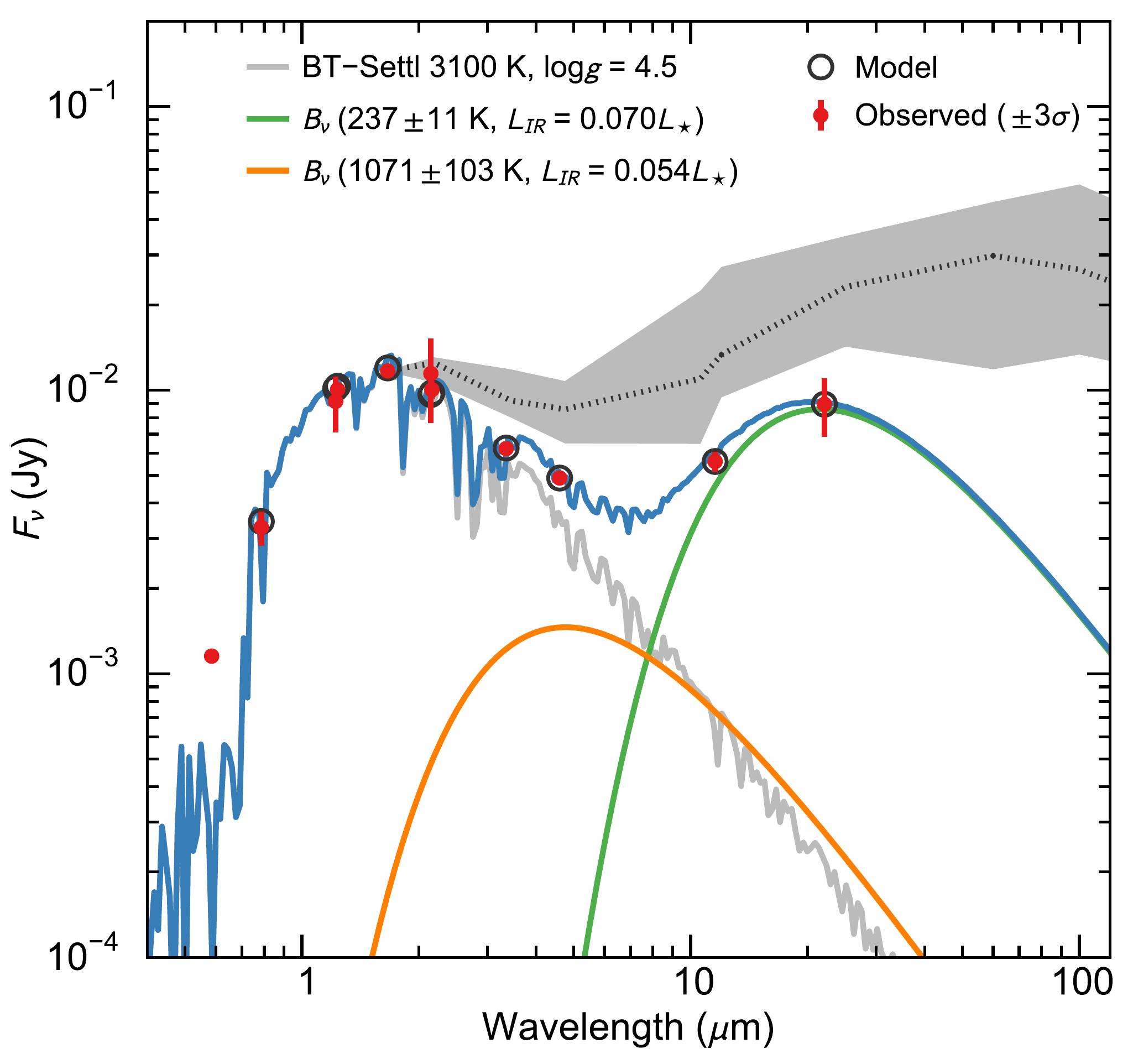}%
   \caption{Spectral energy distribution of \starname, with photometry
     from \emph{Gaia}, DENIS, 2MASS and \emph{WISE} plotted in red
     with their 3$\sigma$ uncertainties. The observed SED (blue line)
     is approximated by the sum of a 3100\,K BT-Settl model and
     blackbodies of 237\,K (0.070\,$L_{\star}$) and 1071\,K
     (0.054\,$L_{\star}$), fit to the DENIS, 2MASS and \emph{WISE}
     data. For comparison, the interquartile range of K5--M2 Class I/II
     Taurus sources from \protect\cite{DAlessio99} is given by the
     shaded region, normalised at 1.6\,$\mu$m.}
   \label{fig:sed}
\end{figure}

The resulting SED and best-fit photosphere model of
$T_{\rm{eff}}\,=\,3100$\,K is shown in Fig.\,\ref{fig:sed}. This
temperature is consistent with the WiFeS M5 spectral type and the
corresponding dwarf temperature (3050\,K) from \citet{Pecaut13}. The
SED $T_{\rm{eff}}$ is $\sim$200\,K hotter than the corresponding
pre-MS temperature (2880\,K) from the scale of \citet{Pecaut13},
however their pre-MS sample of mid-M stars was dominated by younger,
lower surface gravity stars from the $\sim$10--25\,Myr-old $\eta$ Cha,
TW Hya, and $\beta$ Pic groups. Combining these temperature estimates,
we adopt $T_{\rm eff} \simeq$\,3050\,$\pm$\,100\,K for
WISE~J0808--6443.

\begin{table}
\caption[]{Infrared excesses for WISE~J080822.18--644357.3, assuming the intrinsic colours of a pre-MS M5 star from \citet{Pecaut13}. Almost identical excesses are obtained using the main-sequence colours.}
\begin{tabular}{c c c c}
\hline
$E(K_{\rm{s}}-W1)$ & $E(K_{\rm{s}}-W2)$ & $E(K_{\rm{s}}-W3)$ &
$E(K_{\rm{s}}-W4)$\\ (mag) & (mag) & (mag) & (mag)\\
\hline
$0.102\pm0.034$   &   $0.267\pm0.033$   &   $2.252\pm0.039$   &   $4.017\pm0.088$\\
\hline
\end{tabular}
\label{tab:excess}
\end{table}

From the SED fit it is clear that, in addition to the $W3$ and $W4$
bands, there are smaller excesses in both shorter wavelength
\emph{WISE} filters.  This is also apparent when comparing the
observed $K_{\rm s}-$\emph{WISE} colours to the intrinsic 
colours of an M5 star from \citet[see
  Table~\ref{tab:excess}]{Pecaut13}.  We attempted to fit a single
blackbody to the infrared excess ($\sim$470\,K; $L_{\rm
  IR}/L_{\star}=0.095$), but this gave a poor match to the
22\,$\mu$m data. Although not necessarily a physical model, a simple double blackbody disc with a `warm' $\sim$240\,K
outer component having $L_{\rm IR,warm}/L_{\star}=0.070\pm0.015$ and a `hot'
$\sim$1100\,K inner component with $L_{\rm IR,hot}/L_{\star}=0.054\pm0.018$
provides a good fit to the full gamut of \emph{WISE} photometry (see
Fig.\,\ref{fig:sed}). Together, the fractional luminosity of these
components is 0.12\,$L_{\star}$ -- approximately 50 per cent brighter than the
single temperature model of \citet{Silverberg16}. With only four \emph{WISE} fluxes over 3--22\,\micron\ (see discussion in Sec.\,\ref{sec:evo} below), we made no attempt to fit more complex disc models \citep[e.g.][]{Robitaille17}, which suffer from degeneracies due to a large number of free parameters (typically 7--12 for the models above).

Assuming large grains are responsible for the infrared
excess, these blackbody temperatures correspond to emission radii of
$\sim$25\,\Rsun\footnote{\Rsun\, = 695,700\,km = nominal solar radius
following IAU 2015 Resolution B3 \citep{Mamajek15b}.}, (0.115\,au) and
1.2\,\Rsun\, (0.0056\,au), respectively \citep{Backman93}. The hot
dust at the latter radius is likely associated with the gas feeding
the accretion indicated by the H$\alpha$ and \ion{He}{i} emission
lines at the inner edge of the disc. We note that the temperature of the inner component is
comparable to the temperatures where amorphous silicates anneal into
crystalline form, before these sublimate at 1300--1400\,K
\citep{Henning09}.  Also, given the star's size and approximate
density ($\sim$5.6\,g\,cm$^{-3}$), orbiting bodies with densities of
$\sim$1--2\,g\,cm$^{-3}$ would have Roche limits of 1.0--1.3\,\Rsun,
indicating that the dust could be produced
by disrupted planetesimals on trajectories taking them too
close to the star. The cooler dust belt seems
analogous to the `warm dust temperature' debris discs with
characteristic temperatures of $\sim$190\,K\, commonly seen among
stars ranging from B- through K-type, and likely represents
populations of small dust grains sublimating ice from icy
planetesimals \citep{Morales11,Morales16}.

\subsection{Evolutionary status of the disc}\label{sec:evo}

A number of classification schemes have been proposed for the evolutionary stages of circumstellar discs. We adopt the nomenclature of \citet{Espaillat12}, which has also been used by \cite{Luhman12} and \citet{Esplin14} in their studies of Upper Scorpius and Taurus, respectively. In this scheme \emph{full discs} are optically thick at near- and mid-infrared wavelengths and have not been significantly cleared of primordial dust and gas; \emph{pre-transitional} \citep{Espaillat07} and \emph{transitional discs} have large inner gaps or holes, respectively; \emph{evolved discs} (sometimes called anemic or homologously-depleted discs) are becoming optically thin but do not possess large holes or gaps, while \emph{evolved transitional discs} are optically thin and have large holes. All of the above classifications can be considered \emph{primordial discs}, whereas \emph{debris discs} are composed of second-generation dust generated by collisions of planetesimals after the primordial disc has dissipated.

\begin{figure}  
   \centering \includegraphics[width=0.95\linewidth]{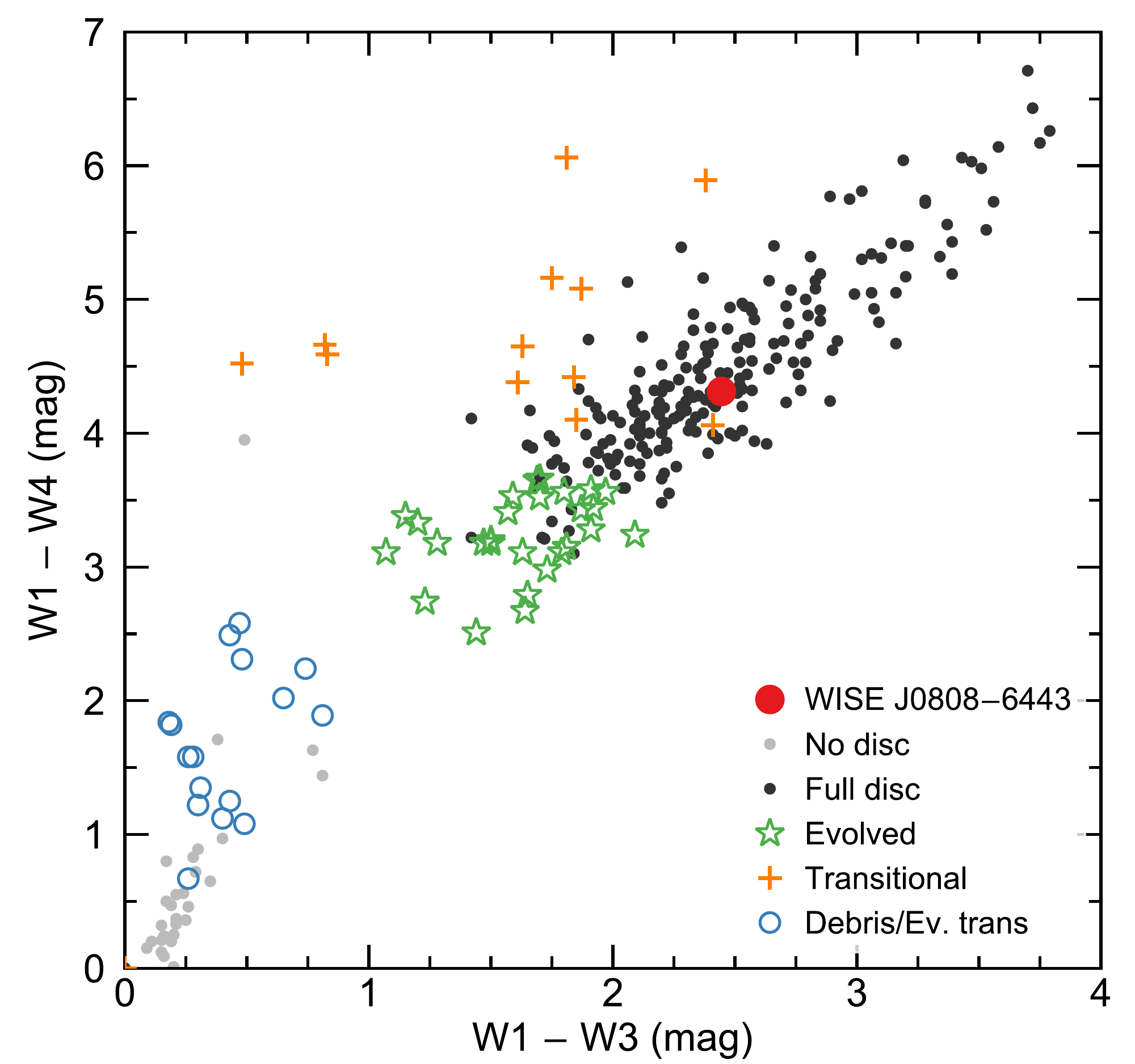}\\
   \centering \includegraphics[width=0.95\linewidth]{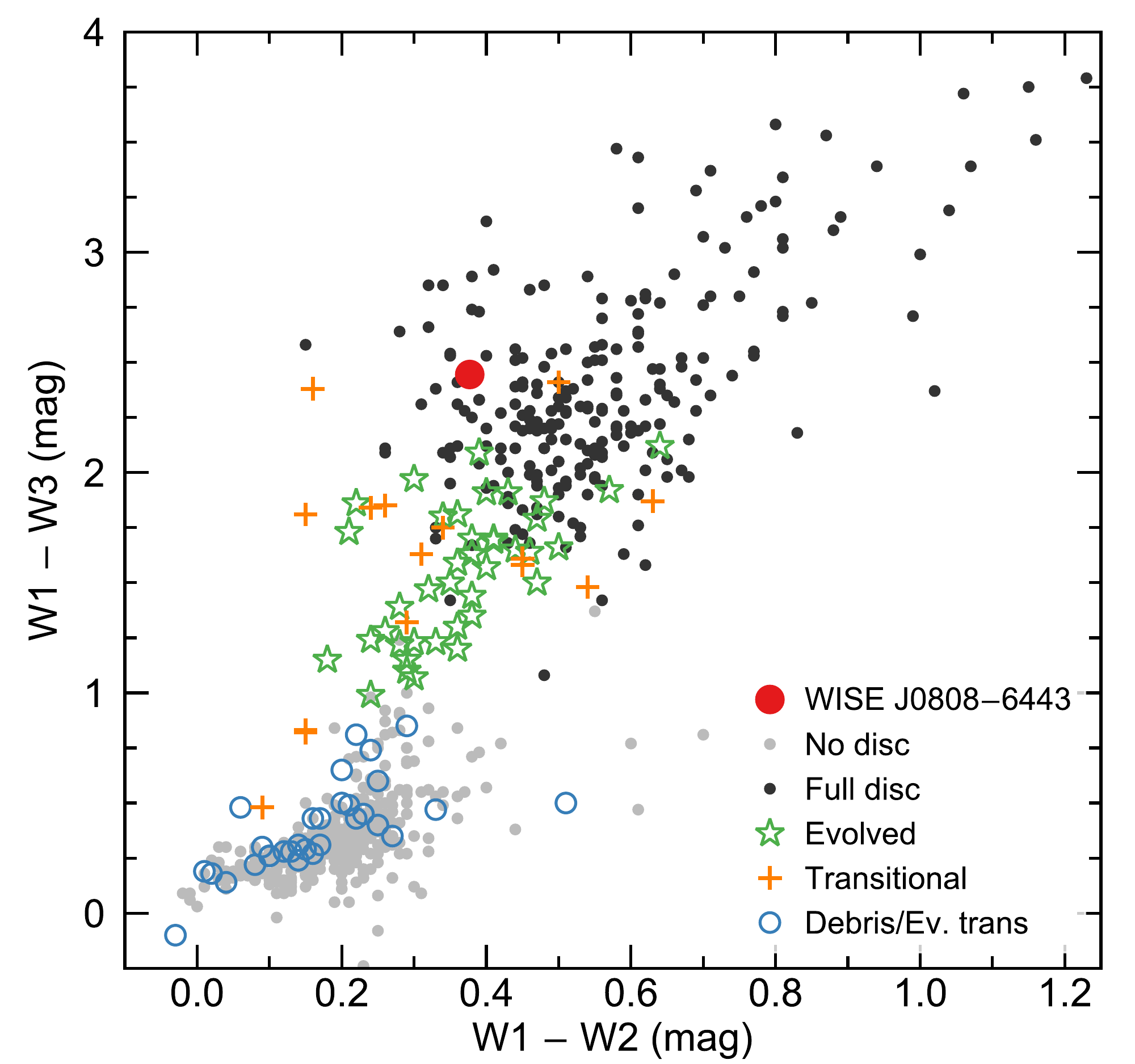}%
   \caption{All\emph{WISE} colour-colour diagrams comparing \starname\ to M dwarf (predominantly M4--M6) members
   of the Taurus and Upper Sco star-forming regions from the studies of \citet{Esplin14} and \citet{Luhman12}, respectively. The evolutionary state of each disc is given by the symbol shape in the legend.} \label{fig:evo}
\end{figure}

These classes are mapped to All\emph{WISE} colours in Figure~\ref{fig:evo}, where we show the classifications of Upper Scorpius and Taurus M~dwarfs from \citet{Luhman12} and \citet{Esplin14}. Moving from red to blue there is a clear evolutionary sequence from full to evolved to debris discs. However, it is important to note that both studies used \emph{Spitzer}/IRAC 8\,\micron\ photometry in addition to \emph{WISE} to separate transitional from full discs, which remain somewhat degenerate in Figure~\ref{fig:evo}.  From its location in this diagram one could classify \starname\ as hosting a full disc without significant clearing of its primordial material and a continuous distribution of dust emitting from the sublimation temperature outwards. However, the SED of the star appears somewhat evolved compared to the median of Class I/II (full disc) sources in Taurus reported by \citet[][dotted line in Fig.\,\ref{fig:sed}]{DAlessio99} and is well-fit by a double blackbody model with a hot inner disc and cooler outer disc. In the absence of flux measurements between $W2$ (4.6\,\micron) and $W3$ (12\,\micron), we cannot definitively distinguish between a full/continuous disc or a (pre-)transitional disc with optically thick inner and outer components separated by an optically thin gap. Irrespective of the final classification, the All\emph{WISE} colours, large fractional luminosity and the detection of ongoing accretion indicate a primordial circumstellar disc and rule out \starname\ hosting a debris disc, as proposed by \citet{Silverberg16}.

\subsection{Mid-infrared variability}

\begin{figure}  
   \centering \includegraphics[width=\linewidth]{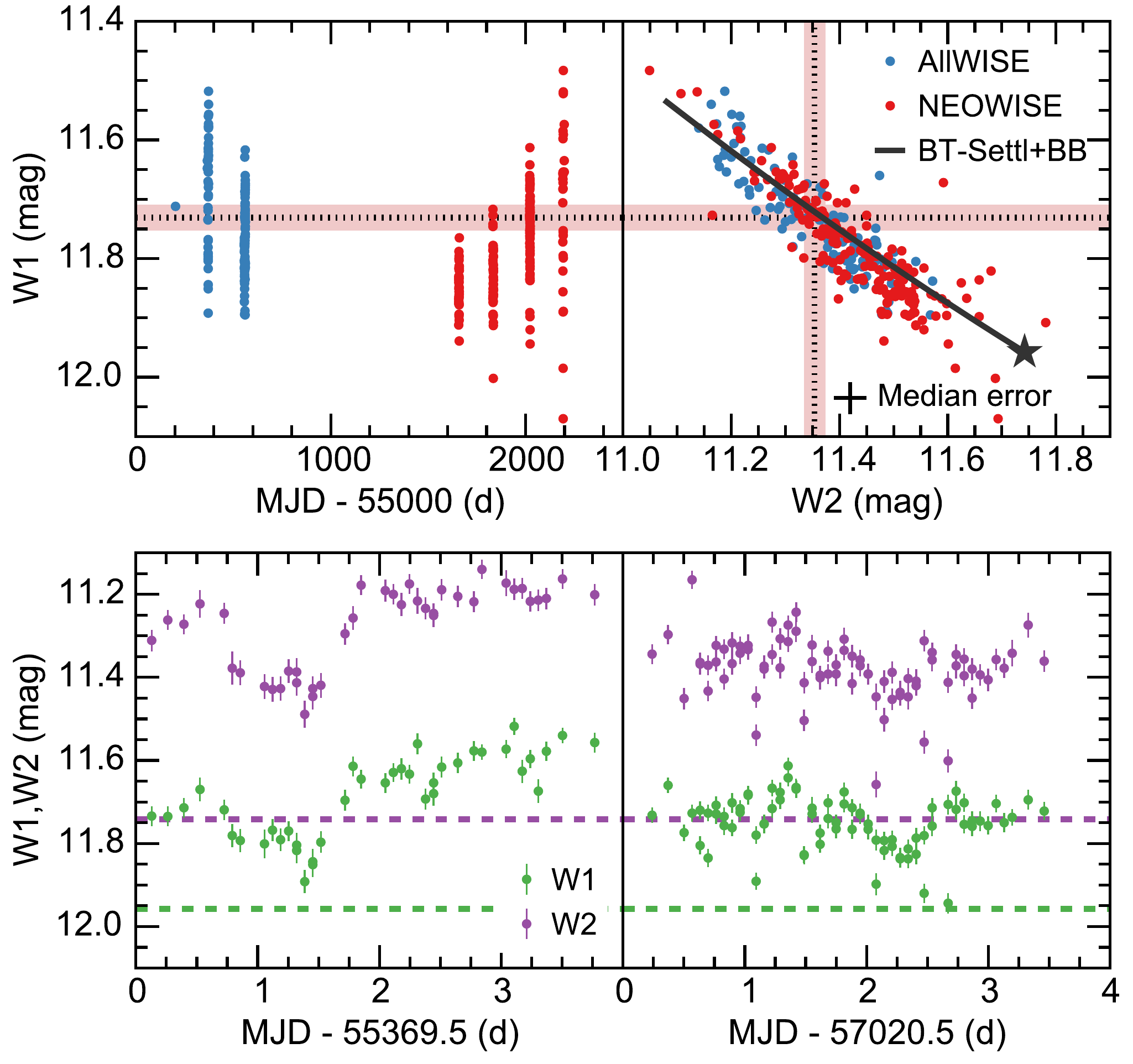}%
   \caption{\emph{Top
   left:} $W1$ individual epoch photometry for \starname\ from both
   the All\emph{WISE} and reactivated NEO\emph{WISE} \citep{Mainzer14}
   missions. \emph{Top right:} Correlation of $W1$ and $W2$
   magnitudes.  The black line shows the expected relation for a
   3100\,K BT-Settl model combined with a 1071\,K blackbody which
   varies in flux from zero (filled star) to 0.11\,$L_{\star}$. The dotted lines 
   are the mean All\emph{WISE} magnitudes used in the SED
   fit ($W1=11.731\pm0.022$\,mag,
   $W2=11.354\pm0.020$\,mag). \emph{Bottom:} $W1$ and $W2$ time series
   for two four~day windows in 2010 and 2014/15. The dashed lines are
   the photospheric contribution calculated from the BT-Settl model
   and correspond to the filled star  in the upper right plot.}  \label{fig:w1w2}
\end{figure}

Both the $W1$ and $W2$ photometry is listed in All\emph{WISE} as
exhibiting strong variability (flag \texttt{var = 9900})\footnote{The $W1$ and $W2$ bands are also flagged as possibly contaminated by diffraction spikes from two bright stars approximately 6 arcmin to the south east and south west (flag \texttt{cc\_flags = dd00}). Visual inspection of the All\emph{WISE} Atlas images shows the photometry is unlikely to be affected.}. This is
illustrated in Fig.\,\ref{fig:w1w2} where we plot $\sim$300 individual epoch
magnitudes for both the All\emph{WISE} and reactivated NEO\emph{WISE}
\citep{Mainzer14} missions. The variation in $W1$ and $W2$ is
extremely well-correlated (Pearson $r\approx0.9$) and much larger than
both the typical uncertainties on a single measurement and the mean
All\emph{WISE} photometry used in the SED fit. We could find no
obvious periodicity in the time series, although the \emph{WISE}
sampling (few day observing blocks every six months) are not conducive
to this. The smaller ($N\approx40$) sample of $W3$ and $W4$ data show
no obvious variability or correlation.  Similar variability has been 
observed with \emph{Spitzer} in younger disc-bearing stars  \citep[e.g NGC 2264;][]{Cody14} and can be attributed to various physical processes depending on the light curve morphology, including variable circumstellar obscuration, accretion instabilities, rotating star spots and structural changes in the disc. In this case the rapid ($\sim$1 day), semi-periodic variation suggests it is driven by changes in the stellar flux modulated by rotation (e.g. cool/hot spots, obscuration) reprocessed into the infrared and not intrinsic changes in the disc, which should occur on longer timescales. The light curve can also be approximated as the luminosity of the inner blackbody varying at fixed temperature (see Fig.\,\ref{fig:w1w2}), perhaps reflecting the amount of dust in a clumpy inner disc.

\begin{figure*}  
   \centering
   \includegraphics[width=0.49\linewidth]{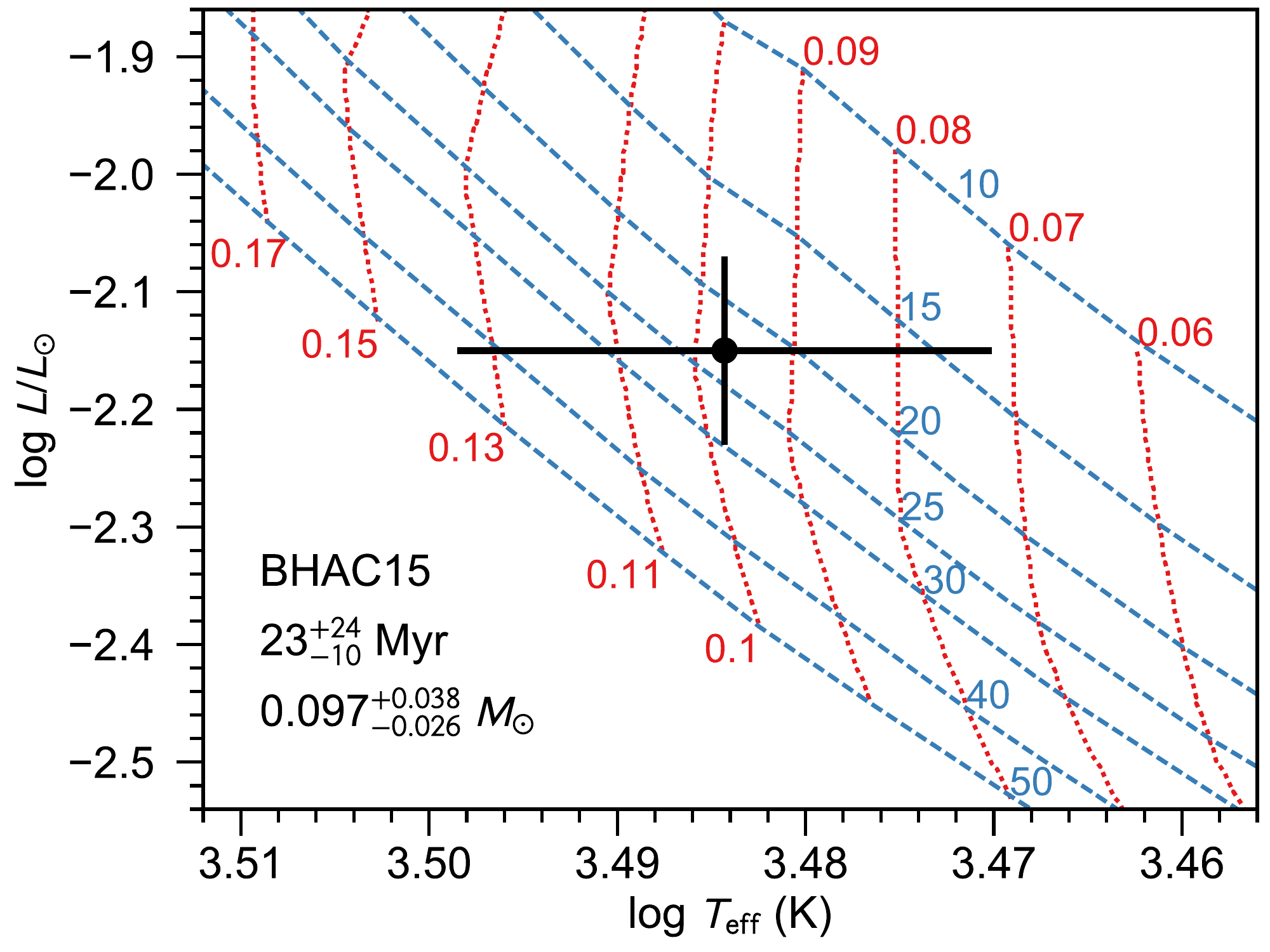} \hfill
   \includegraphics[width=0.49\linewidth]{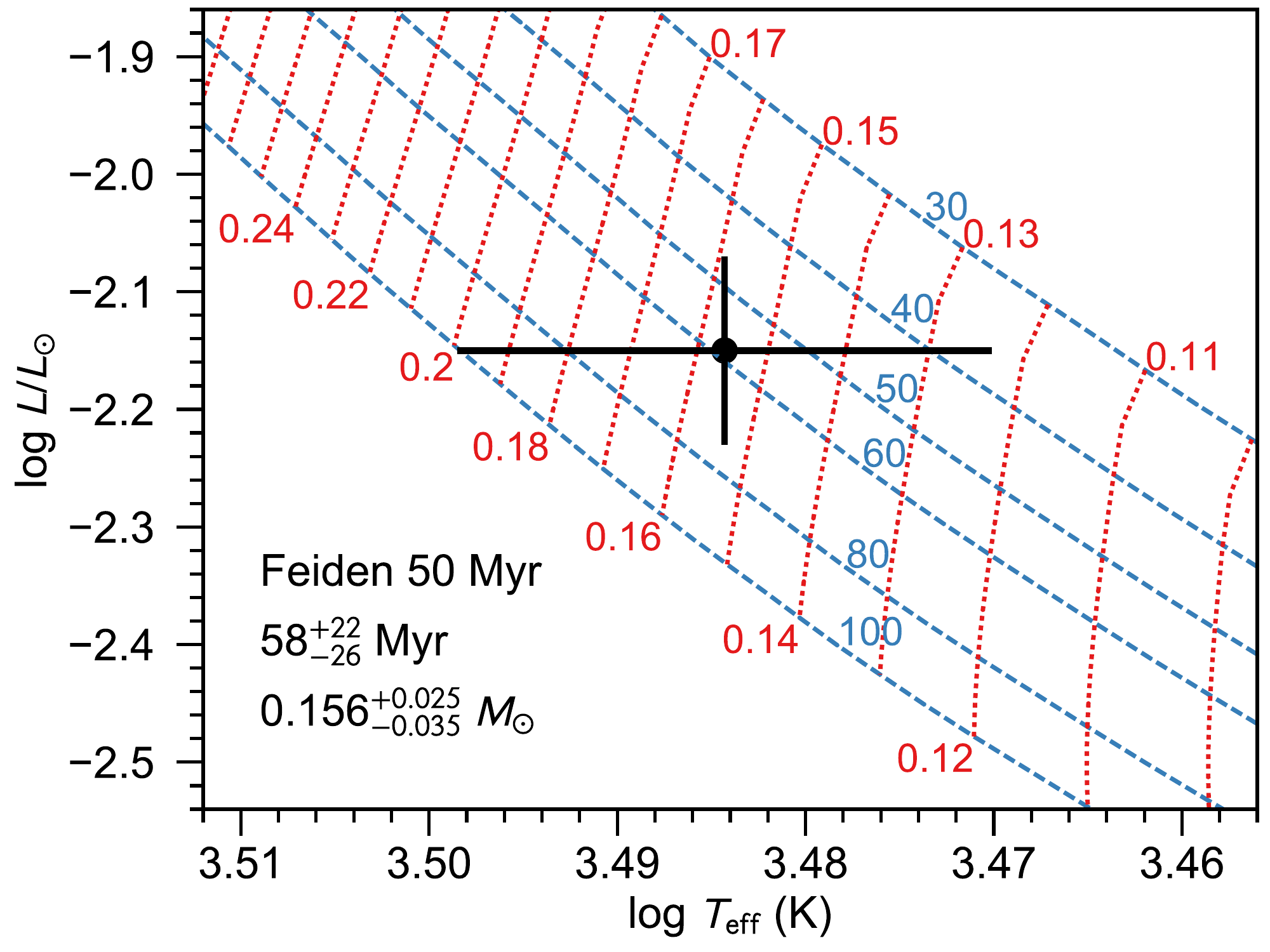}\\%
   \includegraphics[width=0.49\linewidth]{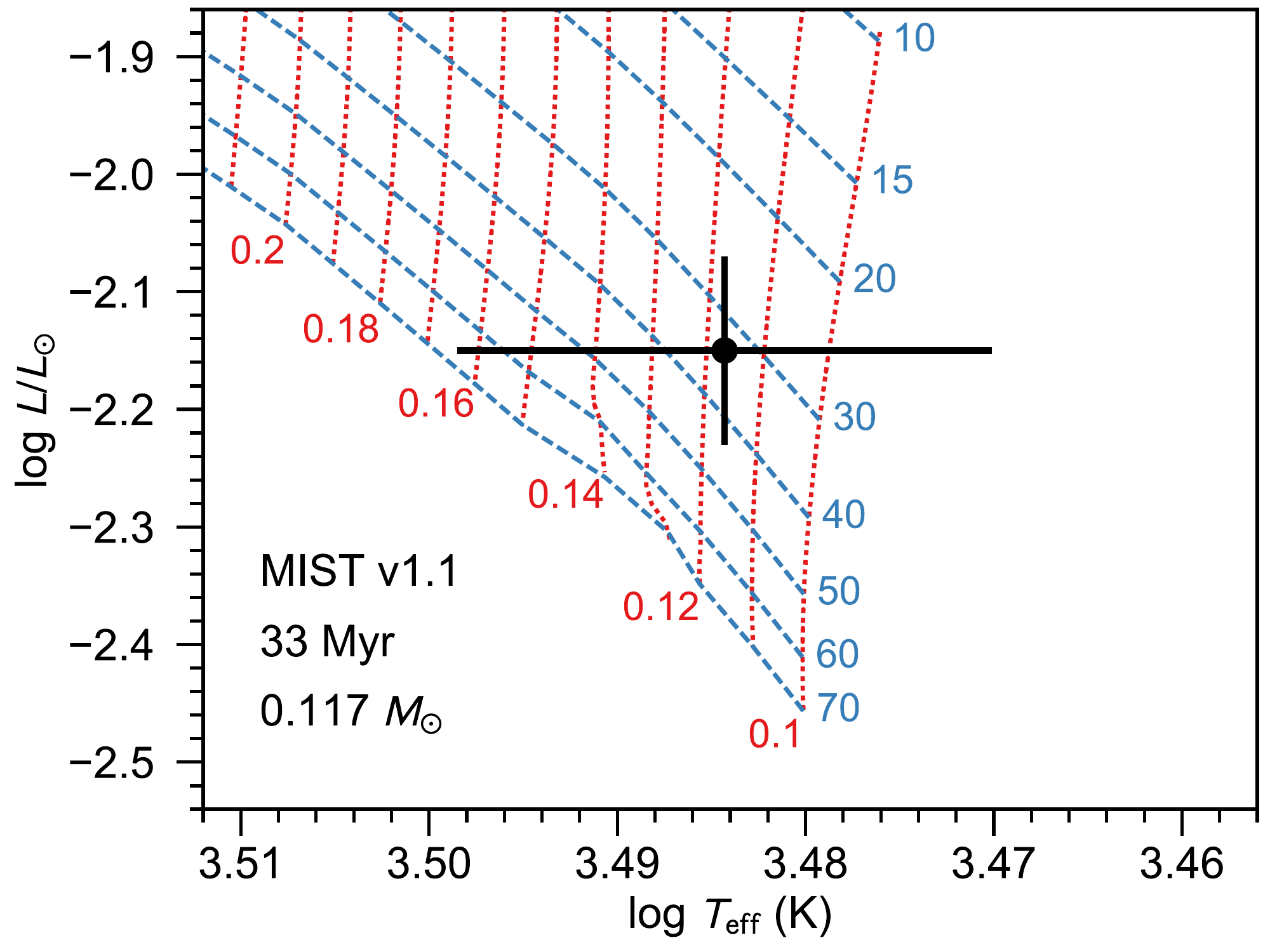} \hfill
   \includegraphics[width=0.49\linewidth]{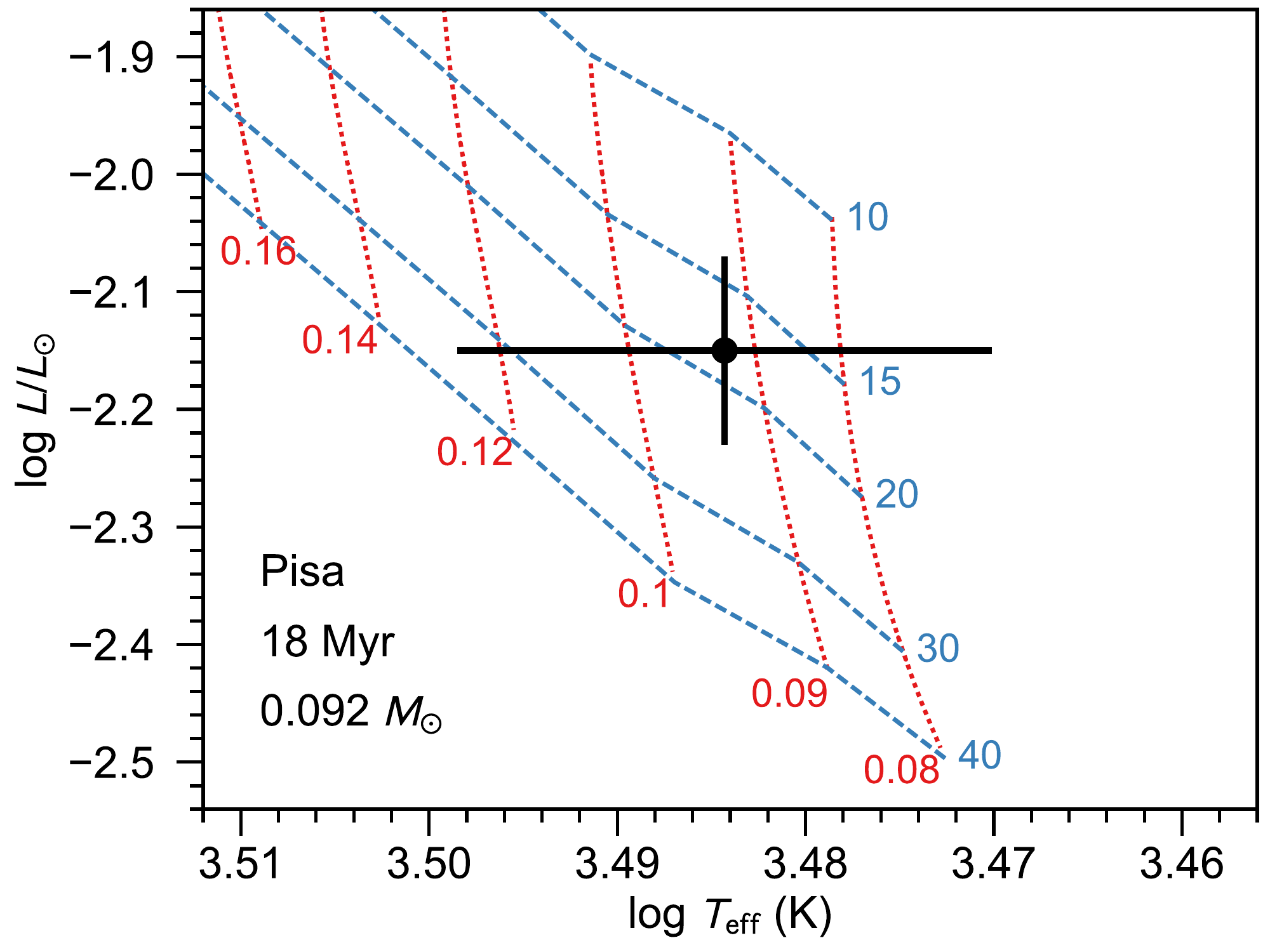}\\%
   \caption{H-R diagram position for \starname, with isochrones and
     evolutionary tracks for solar metallicity pre-MS stars
      from \citet{Baraffe15} (top left), 50 Myr magnetic field equipartition models based on \citet{Feiden16} (top right), MIST v1.1 \citep[bottom left,][]{Dotter16,Choi16} and the Pisa models of \citet{Tognelli11} (bottom right).  Interpolating at the position of \starname\ we estimate ages from these models of 23$^{+24}_{-10}$\,Myr, 58$^{+22}_{-26}$\,Myr, 33\,Myr and 18\,Myr, respectively. Uncertainties on the BHAC15 and Feiden age and mass estimates are the 68 per\,cent confidence intervals and were calculated from a Monte Carlo simulation assuming the quoted uncertainties on $T_{\rm eff}$ and \logl.}
   \label{fig:hrd}
\end{figure*}

\section{Isochronal age of \starname}\label{sec:iso}

Integrating the 3100\,K BT-Settl model atmosphere to 1000\,$\mu$m we
find \logl\,=\,$-2.15\pm0.08$\,dex at a distance of $90\pm9$\,pc, in
agreement with the \logl\,=\,$-2.17\pm0.10$\,dex obtained using the
2MASS $J$-band magnitude and the relation between $T_{\rm{eff}}$ and
the $J$-band bolometric correction given
in \cite{Pecaut13}\footnote{The recent IAU 2015 Resolution B2
regarding the IAU bolometric magnitude scale \citep{Mamajek15a}
suggests an adopted $M_{\rm{bol}, \odot}$ of 4.74\,mag. This differs
from the value adopted by \cite{Pecaut13} by $-0.015$\,mag.  To
conform with the new resolution and place the luminosity
of \starname\, on the 2015 IAU scale, we have modified the $J$-band
bolometric correction from \citeauthor{Pecaut13} accordingly.}. The
estimated effective temperature (3050\,$\pm$\,100\,K) and bolometric
luminosity of the star are consistent
with a stellar radius of $R_{\star}$ = 0.30\,$\pm$\,0.04\,\Rsun. In
Fig.\,\ref{fig:hrd} we plot Hertzsprung-Russell (H-R) diagrams
for \starname\ against the evolutionary tracks of solar
metallicity stars from \citet{Baraffe15} (BHAC15), MESA Isochrones and Stellar Tracks (MIST) v1.1 \citep{Dotter16,Choi16}, the Pisa 
group \citep[][extended down to 0.08\,\Msun]{Tognelli11} and the recent magnetic
models of \citet{Feiden16} \citep[also see][]{Malo14b}.  The latter
were computed assuming the surface gas pressure is in equilibrium with
the magnetic field pressure and adopt an equipartition magnetic field
strength equal to the value at 50 Myr (typically a few kG). This is an
approximation since the surface gas pressure changes as the object
contracts with time, and hence if \starname\ is significantly younger than 50
Myr the magnetic field strengths will be overestimated and for older
ages slightly underestimated.

The BHAC15, MIST and Pisa models yield ages of 23, 33 and 18~Myr, respectively, all with masses $\sim$0.1\,\Msun, while the magnetic models favour an older age ($\sim$60~Myr; similar to the assumed 50~Myr equipartition age) and larger mass (0.15\,\Msun). Bisecting these is the 45$^{+11}_{-7}$\,Myr isochronal age of Carina found by \cite{Bell15} from 12 members and the BHAC15, Pisa, Dartmouth \citep{Dotter08} and PARSEC \citep{Bressan12} models.  Although \starname\ is slightly cooler than the least-massive 
0.1\,\Msun\ Dartmouth evolutionary track, its 1$\sigma$ error ellipse falls inside the grid and we estimate an age of $\sim$23\,Myr. Similarly, we find an age of $\sim$30~Myr using the older but commonly-cited isochrones of \citet{Siess00}. The \starname\ error ellipse does not overlap the PARSEC v1.1 models at all, while the newest 1.2S version implies a much larger age of 100--200~Myr and mass $\sim$0.25\,\Msun. These models were substantially updated for low-mass dwarfs by \cite{Chen14}, who implemented $T-\tau$ relations from the BT-Settl model atmospheres as surface boundary conditions, calibrated to reproduce the observed mass-radius relation of low-mass dwarfs. \cite{Kraus15} have suggested that this may be an over-correction in the pre-MS regime, as expected if the recalibration is primarily due to missing opacities which are more important at higher gravities. From the four sets of models in Fig.\,\ref{fig:hrd} we calculate a mean isochronal age for \starname\ of $33_{-15}^{+25}$~Myr, in agreement with the isochronal age of Carina. Regardless of the exact age adopted and despite recent revisions to young cluster ages \citep{Bell13},  \starname\ appears to be several times older than typically-quoted circumstellar disc and accretion lifetimes (see Introduction).

\begin{table}
\caption[]{Summary of WISE~J080822.18--644357.3 parameters.}
\label{summary}
\begin{tabular}{l l l}
\hline
Parameter & Value & Units\\
\hline
Right Ascension & 122.0924291 & $^{\circ}$ (\emph{Gaia})\\
Declination & $-$64.7325757 & $^{\circ}$ (\emph{Gaia})\\
Spectral type      & M5 & \\
$T_{\rm eff}$     & $3050\pm100$  &   K\\
$\mu_{\alpha}\cos\delta$ & $-12.5\pm2.1$ & \masyr \\
$\mu_{\delta}$ & $+29.4\pm2.5$ & \masyr\\
RV              & $22.7\pm0.5$  &   \kms\\
Distance           & $90\pm9$         & pc\\
EW[\ion{Li}{i}] & $380\pm20$       & m\AA\\
\logl\, & $-2.15\pm0.08$ & dex\\
Radius          & $0.30\pm0.04$ & \Rsun\\
Mass$^{a}$            & 0.16$^{+0.03}_{-0.04}$ & \Msun\\
Age             & 45$^{+11}_{-7}$  (Carina) & Myr\\
& $33_{-15}^{+25}$ (isochronal) & Myr\\
$(U,V,W)$ & $(-10.3, -23.8, -5.2)$ & \kms \\
& $\pm (1.4, 1.1, 1.0)$ & \\
$(X,Y,Z)$ & $(12.1, -85.4, -25.8)$ & pc\\
& $\pm (4.9, 6.3, 4.2)$ &\\
$T_{\rm disc,hot}$  & $1071\pm103$ & K\\
$L_{\rm disc,hot}$  & $0.054\pm0.018$ [0...0.11]$^{b}$ & \Lstar\\
$T_{\rm disc,warm}$ & $237\pm11$ & K\\
$L_{\rm disc,warm}$ & $0.070\pm0.015$ & \Lstar\\
\hline
\end{tabular}
\vspace{1pt}
\begin{flushleft}
  $^{a}$ Mass inferred from HR diagram position, interpolating from the
  evolutionary tracks of \citet{Feiden16}. \\
  $^{b}$ Approximate range from variability observed in $W1$ and $W2$ photometry (see Fig.\,\ref{fig:w1w2}).
\end{flushleft}
\label{tab:summary}
\end{table}

\section{Conclusions}

The parameters of \starname\ determined in this work are summarised
 in Table~\ref{tab:summary}. Based on all available spectroscopy, photometry 
 and astrometry,
we conclude the star is a $\sim$45\,Myr-old M5 ($\sim$0.1\,\Msun)
member of the Carina association at a distance of $\sim$90\,pc and
is actively accreting from a gas-rich circumstellar disc.

While rare, the discovery of an accreting low-mass star in the GAYA
 (Carina, Columba, Tuc-Hor) complex is not
unprecedented.  \citet{Reiners09} describe a Li-rich M7.5 star with
strong, asymmetric H$\alpha$ emission which they attribute to
accretion ($\log \dot{M}_{\rm acc}/M_{\odot}\approx-10.9$). Their
tentative membership assignment to the Tuc-Hor association was
confirmed by \cite{Gagne14}. Recently, \citet{Boucher16} reported
finding two brown dwarfs in Columba and Tuc-Hor which host
circumstellar discs and show signs of accretion (H$\alpha$, Pa$\beta$
emission, respectively). The 0.15\,$L_{\star}$ fractional disc
luminosity of their Tuc-Hor member is similar to the value we find
for \starname\ and is typical of primordial or early pre-transitional
discs. The discovery of these
older, low-mass accretors in nearby moving groups could be an indication that at least
some low-mass stars and brown dwarfs are able to retain their gas
reservoirs \emph{much} longer than previously
recognised. Alternatively, some transient mechanism, for example the tidal disruption of inner gas-giant planets or the collisional grinding and accreting of icy comets, may
be responsible for generating short-lived, gas-rich discs during the
later stages of pre-MS evolution. \citep[e.g.][]{Melis12}

Finally, the existence of a $\sim$45\,Myr-old accreting M dwarf expands the
parameter space for investigations of planet formation around low-mass
stars. Giant planets are observed less frequently in orbit around M dwarfs \citep{Johnson10} and this deficiency is often attributed to the inability of core accretion to build up a large enough mass within the $<$10~Myr lifetime of a typical disc \citep{Laughlin04}.  The presence of long-lived gas in the disc could contribute
multiple important effects to the star's growing planetary system,
such as (1) prolonged accretion of H and He onto the envelopes of
growing proto-planets, (2) prolonged gas-drag which could dampen the
eccentricities of growing planetesimals and planets, (3) a prolonged
epoch of disc torques which may lead to planet migration, or the
formation of planets from matter shepherded by moving secular
resonances \citep[e.g.][]{Raymond08,Ogihara09}.
Finally, the characteristics of the lowest mass pre-MS stars
like \starname\, and their circumstellar material can also provide
important constraints on the conditions that spawn compact systems of
small planets in mean-motion resonances orbiting closely around some
low-mass stars
\citep[e.g. TRAPPIST-1;][]{Gillon16}.

\section*{Acknowledgements}

The authors wish to thank Joao Bento (ANU) for the loan of observing
time to obtain the R3000 spectrum, Grant Kennedy (Cambridge) for
fruitful discussions concerning the circumstellar disc excess, Greg
Feiden (North Georgia) for calculating new magnetic evolutionary
tracks
and the anonymous referee for a prompt and thorough review of the manuscript.
SJM acknowledges support from a University of New South Wales Vice
Chancellor's Fellowship.
EEM acknowledges support from the NASA NExSS program.
Part of this research was carried out at the Jet Propulsion
Laboratory, California Institute of Technology, under a contract with
NASA.
This publication makes use of VOSA, developed under the Spanish
Virtual Observatory project supported from the Spanish MICINN through
grant AyA2011-24052.
This work has made use of data from the European Space Agency (ESA)
mission {\it Gaia} (\url{http://www.cosmos.esa.int/gaia}), processed
by the {\it Gaia} Data Processing and Analysis Consortium (DPAC).
This publication makes use of data products from the \emph{Wide-field Infrared Survey Explorer}, which is a joint project of the University of California, Los Angeles, and the Jet Propulsion Laboratory/California Institute of Technology, and NEOWISE, which is a project of the Jet Propulsion Laboratory/California Institute of Technology. WISE and NEOWISE are funded by NASA.

\bibliographystyle{mnras}
\bibliography{paper}
\bsp
\label{lastpage}
\end{document}